\documentclass[amsmath,amssymb,onecolumn,superscriptaddress]{revtex4}

\usepackage{graphicx}
\usepackage{dcolumn}

\def\be{\begin{equation}}
\def\ee{\end{equation}}
\def\beq{\begin{eqnarray}}
\def\eeq{\end{eqnarray}}

\usepackage{bm}
\usepackage{graphicx}
\usepackage{dcolumn}
\usepackage{amsmath}
\usepackage[latin1]{inputenc}
\usepackage{graphicx, psfrag}
\usepackage{amssymb}
\usepackage[colorlinks=true, citecolor=blue, urlcolor = blue, linkcolor= red, bookmarks=true]{hyperref}
\usepackage{float}
\usepackage{amsmath}
\usepackage{tensor}
\usepackage{amsfonts}
\usepackage{dcolumn}
\usepackage{hyperref}
\usepackage{subfigure}
\usepackage{pgfplots}
\usepackage{epstopdf}
\usepackage{booktabs}
\usepackage{pdfpages}

\begin{document}

\title{ Dark Energy Compact Stars in Extended Teleparallel Gravity}

\author{Allah Ditta}
\email{mradshahid01@gmail.com\textcolor{red}{Corresponding Author}}
\affiliation{Department of Mathematics, School of Science, University of Management and Technology,  Lahore, 54000, Pakistan.}
\affiliation{Research Center of Astrophysics and Cosmology, Khazar University, Baku, AZ1096, 41 Mehseti Street, Azerbaijan}

\author{Xia Tiecheng}
\email{xiatc@shu.edu.cn}
\affiliation{Department of Mathematics, Shanghai University  and Newtouch Center for Mathematics of Shanghai University,  Shanghai,200444, P.R.China.}

\author{G. Mustafa}
\email{gmustafa3828@gmail.com}
\affiliation{Department of Physics, Zhejiang Normal University, Jinhua 321004, People's Republic of China}
\affiliation{Institute of Fundamental and Applied Research, National Research University TIIAME, Kori Niyoziy 39, Tashkent 100000, Uzbekistan}

\author{De\u{g}er Sofuo\u{g}lu}
\email{degers@istanbul.edu.tr}
\affiliation{Department of Physics, Istanbul University, 34134, Vezneciler, Fatih, Istanbul, Turkey}

\author{Asif Mahmood}
\email{ahayat@ksu.edu.sa}
\affiliation{College of Engineering, Chemical Engineering Department, King Saud University Riyadh, Saudi Arabia}

\begin{abstract}
This paper examines dark-energy compact stars under the paradigm of modified Rastall teleparallel gravity. This is the primary analysis of dark energy celestial phenomena under this modified gravitational theory. Utilizing the torsion-based functions, $f(T)$ and $h(T)$, we examined their impacts within a spherically symmetric space-time designated as the inner geometry, while employing the Schwarzschild geometry as the outside space-time. This study examines several features of dark energy in stars, encompassing dark energy pressure components, energy conditions, and equation of state components. Our findings indicate that the detected adverse behavior of certain stellar parameters provided substantial evidence, ensuring the presence of dark energy in celestial configurations. Thorough examinations of energy conditions, pressure profiles, sound speeds,  adiabatic index, gradients, mass function, compactness, and redshift function provide a full evaluation, confirming the viability and authenticity of the analyzed stellar configuration.\\
 \textbf{Keywords}: Anisotropic fluid; Dark energy stars; Modified Rastall Teleparallel Gravity; $f(T)$ modified gravity.
\end{abstract}
\date{\today}
\maketitle
\section{Introduction}\label{sec1}
A revolutionary finding about the cosmos was made in 1998 \cite{riess1998observational}: it was expanding faster than observers had predicted. Scientists became quite curious about what was causing this expansion. Eventually, they discovered that the source of this expansion was dark energy, also referred to as ($\Lambda$) the cosmological constant, which is the negative pressure fluid. However, it disagreed with the energy conditions, which made many scientists question whether the General Theory of Relativity (GR) could deal with it satisfactorily. GR theory explains Local gravity well, which is dependent upon the symmetric and torsion-less Levi-Civita connection \cite{aldrovandi2012teleparallel}. \textcolor{black}{GR effectively elucidates gravitational processes at many scales, from local systems like the Solar System to cosmological structures, contingent upon the assumption of dark energy and DM. Nonetheless, GR encounters difficulties in elucidating the accelerated expansion of the universe and the rotation curves of galaxies without supplementary modifications \cite{berti2015testing,clifton2012modified}. Furthermore, the phenomenological constraints of GR become evident at quantum scales, necessitating a comprehensive theory of quantum gravity \cite{rovelli2004quantum}}. To overcome this weakness, scientists began to think about altering GR. A large number of these modifications aimed at expanding the geometry of GR, among which the most successful were $f(R)$ theories. The Lagrangian function in such hypotheses is written as $R$; the Ricci scalar \cite{sotiriou2010f,bamba2012dark}.

 Because it closely resembles General Relativity (GR), teleparallel gravity, or TEGR, has become increasingly popular. TEGR is a curvature-free theory based on torsion. \textcolor{black}{Although the two theories are formulated very differently, the Teleparallel Equivalent of General Relativity (TEGR) reproduces the identical field equations as GR. TEGR is based on spacetime torsion, with the tetrad field and the Weitzenbock connection as its fundamental elements, whereas GR depends on spacetime curvature defined by the metric tensor and the Levi-Civita connection \cite{aldrovandi2012teleparallel,maluf2013teleparallel}. Furthermore, a fundamental tenet of GR, the equality of inertial and gravitational masses, is not necessary for TEGR \cite{pereira2014teleparallelism}. Despite these distinctions, TEGR is a workable reformulation of GR because the mechanical results of the two theories are identical}. These two ideas, albeit similar, are not understood mathematically in the same way. \textcolor{black}{The $f(T)$ gravity theory is a more generalized variant of TEGR and is closely related to the $f(R)$ theory in terms of procedure. Although the approaches taken by the $f(T)$ and $f(R)$ gravity theories to extend GR are similar, they are fundamentally distinct. While $f(R)$ gravity alters the curvature scalar $R$ in the setting of GR, $f(T)$ gravity alters the teleparallel framework by generalizing the torsion scalar $T$ \cite{bamba2012dark}. However, by adding the boundary term $B$, which connects $T$ and $R$, $f(T,B)$ gravity unites these methods. Curvature-based and torsion-based theories are directly related, as demonstrated by the subclass $f(T,B)=f(-T+B)=f(R)$, which recovers $f(R)$ gravity \cite{bahamonde2019can}}. In contrast to General Relativity, $f(T)$ theory is curvature-free and has non-zero torsion since it is founded on the Weitzenbock connection \cite{aldrovandi2012teleparallel}. Torsion was the foundation for Einstein's original definition of space-time \cite{einstein1925einheitliche}. Unlike the metric function, the tetrad is important in TEGR for setting up the field equations. Selecting the appropriate tetrad is essential to releasing the function ``$f$" from constraints because different tetrads can result in distinct field equations. This framework offers a basis for TEGR modification. \textcolor{black}{A details review about the teleparallel gravity can be found in \cite{bahamonde2023teleparallel}.} \textcolor{black}{In the context of $f(T)$ gravity, Birkhoff's theorem holds true, but the selection of tetrad is crucial to determining the field equations. In order to satisfy the antisymmetric field equations, a diagonal tetrad can be employed, provided that the right spin connection is selected \cite{krvsvsak2015spin}. The equations for $f(T)$ and $T$, however, are not necessarily limited to diagonal tetrads, and non-diagonal tetrads frequently offer more flexibility and convenience when investigating solutions \cite{tamanini2012good}. Due to this reason, we use a non-diagonal tetrad in the current study.} Some good works to review the $f(T)$ gravity are available in literature \cite{bahamonde2023perturbations,bahamonde2018thermodynamics,bahamonde2022black,bahamonde2019photon,bahamonde2023teleparallel}.

The energy-momentum tensor (EMT) conservation law connects space-time geometry and matter in Einstein's General Relativity (GR). But this conservation is only valid under certain circumstances, leaving GR open to modify. While Rastall challenges the standard assumption in GR, $\Theta^{\nu}{}{\mu;\nu}=a_{\mu}$, where $a_\mu$ nullifies in flat spacetime \cite{rastall1972generalization}, the standard assumption in GR is $\Theta^{\nu}{}{\mu;\nu}=0$. Space-time curvature is caused by matter, which causes gravity, and vice versa. Matter is impacted by tidal gravitational forces, which are caused by curvature. Let $\Theta^{\nu}{}{\mu;\nu}=0$ in flat spacetime. With Rastall's adjustment, the non-minimal coupling between geometry and matter is measured by a coupling parameter, $\lambda$. For $\lambda=0$, it reduces to GR. Rastall gravity has been modified to incorporate solutions for unique black holes, such as $\nabla\mu T^{\mu\nu}=\nabla_\nu(\lambda^\prime R)$ \cite{moradpour2017generalization} and $\nabla_\mu T^{\mu\nu}=\lambda \nabla^\nu f(R)$ \cite{lin2019neutral}. In the references \cite{saaidi2020traversable,nazavari2023wormhole}, Rastall-based adjustments are extended to teleparallel gravity. Our focus in this research is teleparallel gravity with Rastall effects \cite{nazavari2023wormhole}.

\textcolor{black}{The modified Rastall theory has garnered considerable interest within the realm of compact star physics owing to its distinctive characteristics. Rastall theory shares all vacuum solutions with GR; however, its non-vacuum solutions exhibit significant differences upon the introduction of the Rastall parameter \(\lambda\). This distinction has generated significant interest and discussion among scientists. Visser \cite{visser2018rastall} argued that Rastall theory is fundamentally equivalent to GR, claiming that Rastall's EMT represents a reconfiguration of the matter sector in GR. Darabi et al. \cite{darabi2018einstein,darabi2018einsteina} contested this assertion, positing that Rastall theory is inherently different from GR. They argued that Rastall's EMT aligns with the conventional definition and, to substantiate their claim, employed a method akin to Visser's to illustrate that \(f(R)\) gravity is not equivalent to GR. Rastall theory simplifies to GR at a particular value of the Rastall parameter \(\lambda\). The analysis conducted by Darabi et al. has been corroborated by the recent findings of Hansraj et al. \cite{hansraj2019impact}, thereby reinforcing the theoretical validity and distinctiveness of Rastall theory.}

It is evident that GR has been effective in analyzing the majority of research projects pertaining to different astrophysical phenomena along with stellar models up to this point. However, the credibility of GR on broad cosmological scales has been criticized in the late time accelerating phase of the Universe, according to various observational data models \cite{perlmutter1998discovery,bahcall1999cosmic,riess1998supernova,tegmark2004ivezi} \cite{riess1998observational,de2000flat,peebles2003cosmological}. A negative pressure, or gravitationally opposed, concealed source of energy known as ``dark energy" (DE) is estimated to account for a large proportion of energy sources, or roughly $70\%$ of total energy, in order to explain the Universe's accelerating tendency \cite{ade2016astron}. Nevertheless, in the cosmology community, there is still disagreement over the precise nature and genesis of DE. Currently, multiple GR alternations have been devised to address this undesirable state of the Universe. Modified theories of gravity (MTG) are the term most generally used to refer to these alternative theories. some modified theories are discussed in the refrences \cite{capozziello2010beyond,capozziello2011extended,nojiri2007introduction,bamba2012dark,koyama2016cosmological,sotiriou2010f,saaidi2020traversable,nazavari2023wormhole}.

 Early research indicates that matter organization in a spherically symmetric framework depends on perfect fluid. The coincidence of the tangential component of pressure and the radial component of the pressure enables the implementation of the isotropic criteria ($p_r=p_t$) on EFEs. For improved comprehension of matter dispersion within the celestial objects, Jeans \cite{jeans192282} proposed in 1922 that peculiar criteria predominate within the innermost regions of stellar bodies, suggesting the participation of an anisotropic factor. Anisotropy is only a measure of the departure from isotropy and is expressed as $\Delta (p_t-p_r)$. The literature \cite{herrera1997local,mak2002exact} has a significant amount of comprehensive information to explore the impact of anisotropy on star formations due to spherical symmetry. Anisotropy in the relativistic star phenomenon results from the presence of several types of fluids, including superfluids, external or magnetic fields, phase transitions, rotary motions, and other fluids.

 Scientists are intrigued by the evolution of heavily dense matter in extreme conditions, viewing compact stars as potential final stages in regular star life cycles. Compact stellar objects like pulsars, characterized by high densities and strong magnetic fields, play a crucial role in astrophysics \cite{lattmir}. At first, these things were thought to be well described by a spherically symmetric matter dispersion in an isotropic fluid, where radial pressure and tangential pressure proved equal. Nonetheless, Jeans argued in 1992 that anisotropy was required to explain matter dispersion in a stellar body's special circumstances \cite{jeans}. Anisotropy, denoted as $\Delta=p_t-p_r$, indicating deviation from isotropic conditions, can result from various factors like superfluids, fluid mixtures, solid cores, phase transitions, viscosity, and the magnetic fields inside the star \cite{jeans,lemaitre1933univers,ruderman1972pulsars,bowers1974anisotropic,herrera1997local,weber1999quark,sokolov1980phase}. Some of the recent literature \cite{ditta2023effect,ditta2023quintessence,ditta2023cloud,bhar2023study,sedrakian2023heavy,maurya2024compact,lohakare2023influence,maurya2023effect,sharif2025analysis,malik2022study,ilyas2021compact,sharif2022charged,salako2018anisotropic,nashed2020stable,capozziello2016mass,capozziello2014constraining,capozziello2015stellar} discuss some acceptable models of stellar configurations. Some more studies enriched with sound knowledge are present in the references \cite{nojiri2011unified,nojiri2017modified,astashenok2022maximum,astashenok2021causal,astashenok2020extended,astashenok2020rotating,astashenok2020supermassive,astashenok2015extreme,odintsov2023inflationary}.

DE is typically seen as an origin of the energy opposing the energy limits such as the Strong Energy Condition (SEC), that is formulated as $\rho + 3p \geq 0$, in contrast to the anisotropic matter, which has tangential direction pressure profile $p_t$ and radial direction pressure profile $p_r$, which is $\rho + p_r + 2p_t \geq 0$. The primary reason for this infraction is that DE has a negative, pressured nature. The equation of state (EoS) parameter in cosmology is explained as a dimensionless parameter that assumes a perfect fluid, is defined as a relation of pressure $p$ to density component $\rho$. Taking into account the range of the EoS parameter $\omega$, several types of DE have been thought of up to this point. The cosmological constant, represented by the conventional symbols $\Lambda$, is the most straightforward DE applicant among them, and it allines well to the EoS $p = -\rho$. In terms of physics, $\Lambda$ is a smooth vacuum and static form of energy.
On the other hand, scalar fields like quintessence or moduli are active variable parameters that can alter over time in space. EoS $p = \omega \rho$ defines the quintessential DE, with a specific range of $-1 < \omega < -1/3$. The DE candidate exhibits Phantom (or ghost)-like DE when $\omega < -1$. The possibility of DE as a black hole substitute a core element of a compact relativistic object opens up new avenues for cosmological inquiry. Gravastars absence of an event horizon and center singularity is a crucial characteristic. A few research studies on gravastars can be reviewed to learn more about the solutions and physical properties of gravastars in relation to GR and MTGs \cite{usmani2011charged,rahaman20122+,das2017gravastars,ghosh2017charged,sengupta2020gravastar}. Subsequently, some writers hypothesized a compact relativistic object famed as the ``dark energy star" that is composed of a combination of non-interacting DE and regular matter based on the stable physical structure of gravastars. Some recent literature on the study of DE stars is available \cite{pretel2023radial,astashenok2023compact,bhar2023dark,saklany2023compact,ditta2024dark,das2024acceptable,ditta2024impact} and so on. \textcolor{black}{Some other studies on dark energy are available in the literature; some of this literature may be discussed as: equation of state for dark energy have been discusses in $f(T)$ theory of gravity in ref. \cite{bamba2011equation}, the general framework for reconstructing the $f(T)$ models for dynamical dark energy backgrounds have beed studied in ref. \cite{dent2011f}, holographic dark energy by reconstructing the $f(T)$ gravity have been developed in ref. \cite{hamani2012reconstruction,chattopadhyay2013reconstruction},  the generalized way of constructing effective dark energy models is presented in ref. \cite{astashenok2014effective}.}

{\color{black} The investigation of dark energy stars is vital as they present a theoretical alternative to conventional compact objects such as neutron stars and black holes, potentially connecting local astrophysical events to the cosmic acceleration attributed to dark energy. Rastall Teleparallel gravity integrates the non-conserved energy-momentum framework of Rastall gravity with the torsion-based geometry of teleparallelism, introducing additional parameters such as the Rastall parameter $\lambda$ and the torsion scalar. The features facilitate the modeling of dark energy stars through modified equilibrium conditions and new mass-radius relationships, offering insights that extend beyond General Relativity (GR). This framework effectively addresses cosmological challenges, including the dark energy problem and the cosmological constant issue, while providing new insights into compact objects influenced by exotic matter or negative-pressure energy densities. The application of Rastall Teleparallel gravity to dark energy stars allows researchers to investigate the interaction between local astrophysical systems and the overarching dynamics of modified gravity theories.}

The following plan will guide us as we move forward with the next study stage: We shall review the principles of MRTG and evaluate the FEs with an off-diagonal tetrad in Section-\ref{sec2}. Furthermore, we use the Karmarkar condition to find the generalized solutions in this section. We shall compare the inner and outer geometries in Section-\ref{sec3} to ascertain the constant parameters we used in our stellar modeling. After that, we will explain our findings in Section-\ref{sec4} and conclude in Section-\ref{sec5}.

\section{Basic formalism of MODIFIED Rastall TELEPARALLEL GRAVITY(MTRG) and our generalized solutions}\label{sec2}
A basic knowledge of the internal functioning of stars is essential to comprehending their structure. Spherically symmetric space-time is a useful model in this situation. This space-time has consistent features in all directions, making it amenable to mathematical representation using a variety of models, including the Schwarzschild metric. Scientists may learn a great deal about the behavior and formation of stars by analyzing spherically symmetric space-times. These discoveries have important ramifications for a variety of fields in astronomy. This spherically symmetric space-times is given as:
{\color{black}
\begin{equation}\label{1}
ds^{2}=-e^{\nu(r)}dt^2+e^{\lambda(r)}dt^2+r^2sin^{2}\theta d\phi^{2}+r^2d\theta^{2},
\end{equation}
where $e^{\nu(r)}$ and $e^{\lambda(r)}$ are spherically symmetric spacetime potential components. Here, in this manuscript we derive the field equations by using the above-given spherically symmetric spacetime, while this setup is also valid for the FRLW metric. The Minkowski metric $\eta_{i j}=\operatorname{diag}(-1,1,1,1)$ and the tetrad fields $e^{i}{}_{\mu}$ (where as inverse tetrad $E_{i}{}^{\mu}$) can be used to describe the metric tensor $g_{\mu \nu}$ described on a manifold:
\begin{equation}\label{2}
    g_{\mu \nu}=\eta_{i j} e^{i}{}_{\mu} e^{j}{}_{\nu}
\end{equation}
wherein the Latin letter $(i, j, \ldots=0,1,2,3)$ provides tangent space indices and the Greek alphabet $(\mu, \nu, \ldots=0,1,2,3)$ provides space-time indices. In the realm of mathematics, the Weitzenbock connection can be summed up as:
\begin{equation}\label{3}
    \Gamma^{\alpha}{}_{\mu \nu}=E_{i}{}^{\alpha} \partial_\nu e^{i}{}_{\mu}=-e^{i}{}_\mu \partial_\nu E_{i}{}^{\alpha}.
\end{equation}
The teleparallel hypothesis uses a particular kind of link with zero curvature but non-zero torsion. The definition of the torsion tensor, which can be expressed as follows, depends heavily on these connections:
\begin{equation}\label{4}
    T^\sigma{}_{\mu \nu} \equiv \Gamma^\sigma{}_{\mu \nu}-\Gamma^\sigma{}_{\mu \nu}=E_{i}{}^\sigma\left(\partial_\mu e^{i}{}_\nu-\partial_\nu e^{i}{}_\mu\right).
\end{equation}
The subsequent relation connects the Levi-Civita connection denoted as $\bar{\Gamma}^{\sigma}{}_{\mu \nu}$, to the Weitzenbock connection:
\begin{equation}\label{5}
    \bar{\Gamma}^{\sigma}{}_{\mu \nu}=\Gamma^{\sigma}{}_{\mu \nu}-K^\sigma{}_{\mu v},
\end{equation}
in this case, the contorsion tensor, denoted by $K^{\sigma}{ }_{\mu \nu}$, is provided by:
\begin{equation}\label{6}
   K^{\sigma}{ }_{\mu \nu}=\frac{1}{2}\left(T_{\mu}{}^{\sigma}{}_\nu+T_{\nu}{}^\sigma{}_\mu-T^{\sigma}{}_{\mu \nu}\right)+\frac{1}{6} \epsilon^{\sigma}{}_{\mu \nu \rho} A^{\rho}.
\end{equation}
where $A^{\rho}$ is the pseudo-vector of axial torsion, which is obtained by contracting the torsion tensor with the Levi-Civita symbol $A^{\rho} = \epsilon^{\rho \mu \nu \sigma} T_{\mu \nu \sigma}$, where $\epsilon^{\rho \mu \nu \sigma}$ is the completely antisymmetric Levi-Civita symbol. The torsion scalar expression can be understood as:
\begin{equation}\label{7}
    T=S^{\sigma \mu \nu} T_{\sigma \mu \nu}
\end{equation}
When the super-potential $S^{\sigma \mu \nu}$ equation is provided by:
\begin{equation}\label{8}
    S^{\sigma \mu \nu}=-S^{\sigma \nu \mu}=\frac{1}{2}\left(K^{\mu \nu \sigma}-g^{\sigma \nu} T^{\alpha \mu}{}_{\alpha}+g^{\sigma \mu} T^{\alpha \nu}{}_{\alpha}\right) .
\end{equation}
According to \cite{nazavari2023wormhole}, the amended teleparallel gravity action is as follows:
\begin{equation}\label{9}
    S=S_G+S_m=\frac{1}{4 \kappa} \int e f(T) d^4 x+\int e L_m d^4 x,
\end{equation}
wherein $e$ is the representation of the tetrad field $e^a{_F}$ determinant, $\kappa=4\pi G$ represents the geometric unit, and the function $f(T)$ is torsion dependent function. One can recover the TEGR if $f(T)=-T$ is replaced in Eq. (\ref{9}). Moreover, the matter Lagrangian is notioned by $L_m$. The associated field equation can be obtained by calculating the action's variation with respect to the tetrad field:
\begin{equation}\label{10}
    S_i^{\mu \nu} f_{T T} \partial_\mu T+e^{-1} \partial_\mu\left(e S_i^{\mu \nu}\right) f_T-T^\alpha{ }_{\mu i} S_\sigma^{\nu \mu} f_T \\
-\frac{1}{4} E_i{}^\nu f=-\xi \Theta^\nu_{u, \text{matter}},
\end{equation}
while the normal EMT of the ideal fluid is denoted by $ \Theta^\nu_{u, \text{matter}}$,and $\xi$ represents gravitational parameter.
 Fluid's total EMT, which describes the compact star's core, is expressed by:
\begin{equation} \label{17}
\Theta^\nu_{u, \text{matter}}=-\frac{2}{\sqrt{-g}} \frac{\delta\left(\sqrt{-g} \mathcal{L}_m\right)}{\delta g^{\mu \nu}}.
\end{equation}
It is demonstrated that:
\begin{equation}\label{11}
    \left(S_i^{\mu v} f_{T T} \partial_\mu T+e^{-1} \partial_\mu\left(eS_{i}{ }^{\mu \nu}\right) f_T-T^\sigma{ }_{\mu i} S_\sigma^{\nu \mu} f_T\right. \\
\left.-\frac{1}{4} E_i{}^\nu f\right)_{; v}=0,
\end{equation}
In which the covariant derivative under the Levi-Civita connection is indicated by a semicolon:
\begin{equation}\label{12}
    V^\mu{}_{; \nu}=\partial_\nu V^\mu+\left(\Gamma^\mu{}_{\lambda \nu}-K^\mu{}_{\lambda \nu}\right) V^\lambda,
\end{equation}
for any vector $V^{\mu}$ in space-time. The EMT's covariant derivative likewise vanishes because of the calculation above:
\begin{equation}\label{13}
    \Theta^\nu_{u, \text{matter};\nu} = 0.
\end{equation}
The law of conservation for EMT is identical to our revised teleparallel gravity theory and Einstein's theory. But Rastall's new equation, $T^{\mu}{\nu;\mu}=\lambda R{,\nu}$, called into question Einstein's theory's conservation equation. This formula indicates a relationship between matter and geometry and provides a modified field equation, implying an interesting interaction between the two. We assume a similar thing in our MRTG theory, which is motivated by Rastall's idea. By connecting geometry and matter via the scalar torsion of geometry, we create a relation in which the divergence of the torsion scalar corresponds to the divergence of the EMT $\Theta^\nu_u = \Theta^\nu_{u, \text{matter}} - \lambda \delta^\nu_u h(T)$:
\begin{equation}\label{14}
 \textcolor{black}{  \Theta^{\nu}{}_{u;\nu} = (\Theta^\nu_{u, \text{matter}} - \lambda \delta^\nu_u h(T))_{;\nu},}
\end{equation}
here $h(T)$ represents the analytical function based on torsion and $\lambda$ is an arbitrary real valued constant. Next is the rewrite of the field Eq. (\ref{11}):

\begin{equation}\label{15a}
\textcolor{black}{S_u{}^{\mu \nu} f_{TT} \partial_\mu T + e^{-1} \partial_\mu(S_u{}^{\mu \nu}) f_T - T^\sigma{}_{\mu u} S_\sigma{}^{\nu \mu} f_T - \frac{1}{4} E_u{}^\nu f
= -\xi \big[\Theta^\nu_{u,\text{matter}} - \lambda \delta_\mu^\nu h(T)\big],}
\end{equation}
where the interaction term \( \delta_\mu^\nu h(T) \) ensures consistency with the modified conservation law:

\begin{equation}\label{15}
   \textcolor{black}{ S_u{}^{\mu \nu} f_{TT} \partial_\mu T + e^{-1} \partial_\mu(S_u{}^{\mu \nu}) f_T - T^\sigma{}_{\mu u} S_\sigma{}^{\nu \mu} f_T - \frac{1}{4} E_u{}^\nu f-\delta_\mu^\nu \xi \lambda h(T)=-\xi\Theta^\nu_{u,\text{matter}}}.
\end{equation}
The value of the gravitational parameter in Rastall's theory, $\xi$, is represented as follows in relation to Newtonian gravity:
\begin{equation}\label{16}
    \xi=\frac{4 \gamma-1}{6 \gamma-1} \xi_G,
\end{equation}
where the Einstein coupling constant $\xi_C=4 \pi G$ is represented by the notions $\gamma=\lambda \xi$ and $\xi_G$. Energy momentum tensor for anisotropic matter is given as:
\begin{eqnarray}
    \Theta^\nu_{u,\text{matter}}=\left(\rho+p_t\right) U_u U^\nu-p_t \delta_u^\nu+\left(p_r-p_t\right) v_u v^\nu,
\end{eqnarray}

where $U_u$ denotes the time-like four-velocity vector and $V_u$ represents the unitary space-like vector in the radial direction. The equation $U_0 U^0 = -v_1 v^1 = 1$ is satisfied. Additionally, $\rho$ represents the energy density, whereas $p_r$ and $p_t$ denote the radial and transverse pressures, respectively.
}
Assuming for the moment that the inner matter component is composed of a mixture of two non-interacting fluids, such as isotropic regular matter and anisotropic type unidentified matter (referred to as DE). Thus, the subsequent tensor may be utilized for expressing all energy sources together \cite{harko2011two,rahmansyah2021tov}.
\begin{equation}
\Theta_{\mu \nu}^{eff}=\underbrace{\left(\rho^{D}+p_t^{D}\right) W_\mu W_\nu+p_t^{D} g_{\mu \nu}+\left(p_r^{D}-p_t^{D}\right) V_\mu V_\nu}_{\text {dark energy }}+\underbrace{(\rho+p) U_\mu U_\nu+p g_{\mu \nu}}_{\text {ordinary matter }} .
\end{equation}

In this case, the DE applicant's energy density, radial pressure, and tangential pressure are denoted by the appropriate symbols $\rho^{D},\;p_r^{D}$, and $p_t^{D}$, while the energy density and pressure component of the normal matter is represented by $\rho$ and $p$, respectively. Consequently, the mixed EMT $\Theta_{v}$ non-vanishing component can be expressed in the following form:
\begin{eqnarray}
 \rho_{eff}=\Theta_0^0&=&\rho+\rho^{D}, \\
 p_{reff}=\Theta_1^1&=&-\left(p+p_r^D\right), \\
 p_{teff}=\Theta_2^2&=&\Theta_3^3=-\left(p+p_t^D\right), \\
 \Theta_0^1&=&\Theta_1^0=0 .
\end{eqnarray}

The teleparallel technique in GR includes tetrad fields ($e^i_\mu$) representing coordinates and frames through Greek and Latin indices, respectively. The combination of these indices forms the tetrad matrix, satisfying $e^\mu_i e^i_\nu = \delta^\mu_\nu$ and $e^\mu_i e_\mu^j = \delta^j_i$. This technique extends the manifold to include torsion alongside curvature, with the Riemannian curvature tensor expected to be zero. Torsion-free geometry or tetrad-induced torsion can explain gravity. Tamanini and Bohmer \cite{tamanini2012good} discussed the concept of a ``good tetrad" for studying a wider aspect of cosmologies in $f(T)$ theory. Bohmer et al. \cite{boehmer2011existence} investigated a relativistic system of compact stars in $f(T)$ gravity, favoring off-diagonal tetrad for spherically symmetric solutions due to issues with the diagonal tetrad. Numerous studies on spherically symmetric space-time within $f(T)$ modified gravity are available in the literature, including references \cite{li2011f,ruggiero2015weak,bahamonde2019photon,hohmann2019modified,krvsvsak2019teleparallel,nashed2019rotating}, among others.

\textcolor{black}{Tetrads in both GR and $f(T)$ theories serve as local reference frames that denote the state of an observer in spacetime, with the temporal component aligned with the observer's worldline \cite{misner1973freeman,wald2010general}. In $f(T)$ gravity, the selection of tetrads is complex because of the emergence of additional degrees of freedom and the absence of complete local Lorentz invariance. The distinction requires the careful selection of appropriate tetrads that adhere to spacetime symmetries and ensure consistent dynamics, thereby circumventing the issues linked to inappropriate tetrads \cite{krvsvsak2015spin,golovnev2017covariance}. The selection of off-diagonal tetrads guarantees compatibility with the $f(T)$ framework, aligns with the symmetries of the relevant spacetime, and eliminates ambiguities in the resolution of the field equations, in accordance with established methodologies in the literature \cite{bahamonde2023teleparallel}.}

In this work, to evaluate field equations, we will use the off-diagonal tetrad matrix provided in \cite{boehmer2011existence,tamanini2012good,daouda2012anisotropic}:
\begin{equation}\label{20}
    e^a{ }_\mu=\left(\begin{array}{cccc}
e^{\nu / 2} & 0 & 0 & 0 \\
0 & e^{\lambda / 2} \sin \theta \cos \phi & r \cos \theta \cos \phi & -r \sin \theta \sin \phi \\
0 & e^{\lambda / 2} \sin \theta \sin \phi & r \cos \theta \sin \phi & r \sin \theta \cos \phi \\
0 & e^{\lambda / 2} \cos \theta & -r \sin \theta & 0
\end{array}\right) \text {. }
\end{equation}
in which $g_{rr}=e^{\lambda(r)}$ and $g_{tt}=e^{\nu(r)}$ . In the tetrad field above, $e=det\left(e^a{ }_\mu\right)=r^2 \sin \theta e^{(\nu+\lambda)/2}$ is the determimant. Using Eqs. (\ref{4}), (\ref{7}) and (\ref{8}), the torsion scalar takes the follows form:
\begin{equation}\label{21}
    T(r)=\frac{2 e^{-\lambda}}{r^2}\left(e^\frac{\lambda}{2}-1\right)\left(e^\frac{\lambda}{2}-1-r \nu^{\prime}\right).
\end{equation}
After putting Eqs. (\ref{15}, \ref{17}, and \ref{20}) in the given field Eq. (\ref{1}), the EMT's non-zero elements can be defined given as the following:
\begin{eqnarray}
 \xi \left(\rho+\rho^D\right)&= & \frac{e^{-\lambda / 2}}{r}\left(1-e^{-\lambda / 2}\right) f_T^{\prime}-\left(\frac{T}{4}-\frac{1}{2 r^2}\right) f_T
 +\frac{e^{-\lambda}}{2 r^2}\left(r \lambda^{\prime}-1\right) f_T+\frac{f}{4}+\gamma h(T),\label{25aa} \\
\xi \left(p+p_r^D\right)&= & {\left[\frac{T}{4}-\frac{1}{2 r^2}+\frac{e^{-\lambda}}{2 r^2}\left(1+r \nu^{\prime}\right)\right] f_T-\frac{f}{4}-\gamma h(T)},\label{26aa} \\
\xi \left(p+p_t^D\right)&= & \frac{e^{-\lambda}}{2}\left(\frac{\nu^{\prime}}{2}+\frac{1}{r}-\frac{e^{\lambda / 2}}{r}\right) f_T^{\prime}-\frac{f}{4}-\gamma h(T)+  f_T\Big[\frac{T}{4}+\frac{e^{-\lambda}}{2 r}\Big[\left(\frac{1}{2}+\frac{r \nu^{\prime}}{4}\right)\left(\nu^{\prime}-\lambda^{\prime}\right)\nonumber\\&+&\frac{r \nu^{\prime \prime}}{2}\Big]\Big].\label{27aa}
\end{eqnarray}
The conduct of the results is notably affected by the existence of the term $\gamma h(T)$ connected with the Rastall"s coefficient $\gamma$, and the coefficient $\xi$ in FEs, which may change the energy conditions.

A variety of suppositions for the available options in literature for the functions $f(T)$ and $h(T)$ must be taken into account to generate solutions for compact objects in literature. We use the off-diagonal tetrad in our investigation, which increases the physicality of the analysis without adding any limitations \cite{tamanini2012good,boehmer2011existence} on functions that rely on torsion, that is, $f(T)$ and $h(T)$. In pursuit of broader solutions, we select non-linear models such as the power law form of the $f(T)$ function \cite{bamba2012dark,zubair2016analytic} and the logarithmic model for the $h(T)$ function \cite{ditta2023effect}:
\begin{eqnarray}
    f(T)&=&\beta  T^n,\label{28aa}\\
    h(T)&=&\psi  \log \left(\phi  T^{\chi }\right),\label{29aa}
\end{eqnarray}
wherein $\beta,\;n,\;\psi,\;\phi\;\&\;\chi$ are all arbitrary real constants. The values of constants are chosen such that these values produce an acceptable class of results. Like, $\beta\;\&\;n$ are constants that differentiate the function from being simple teleparallel. \textcolor{black}{The model \( f(T) = \beta T^n \) is developed to investigate deviations from Teleparallel Equivalent of General Relativity (TEGR), with \( \beta \neq 1 \) and \( n \neq 1 \) indicating alterations to teleparallel gravity. When \( \beta = n = 1 \), the model simplifies to \( f(T) = T \), thereby recovering the equivalence of TEGR and GR. To ensure consistency with General Relativity at small scales, we establish \( n = 1 \), thereby constraining deviations to large-scale behavior. The selection of \( \beta \neq 1 \) distinguishes the model from TEGR, confirming its status as an extension of teleparallel gravity. In the expression \( h(T) = \psi \log(\phi T^\chi) \), the constraints \( \phi > 0 \), \( \psi \neq 0 \), and \( \chi \neq 0 \) are essential for maintaining a well-defined logarithmic form, which is significant for torsion-based modifications. Parameters \( \gamma \neq \frac{1}{4}, \frac{1}{6} \) prevent singularities arising from Rastall corrections. These selections collectively guarantee that the models maintain mathematical consistency, possess physical significance, and converge to General Relativity under suitable conditions.} {\color{black}{The selected ansatz \( h(T) = \psi \log \left( \phi T^{\chi} \right) \) is chosen based on assumption to comply with the results and may not satisfy the EMT conservation requirement in Rastall teleparallel gravity in the general context. It may fulfill the conservation law of EMT based on some constraints, but generally it does not hold.}}

Eisenhart class-I $(n+1)$-dimensional space $V^{n+1}$ can be embedded into a $(n+2)$-dimensional Euclidean space $E^{n+2}$, as shown by Eisenhart \cite{eisenhart1997riemannian}, as long as a symmetric tensor exists satisfying the Gauss-Codazzi equations:

\qquad \qquad $R_{stpq}=2ea_{s[p{}a_{q}]t}$ \qquad and \qquad $a_{s[t;p]}-\Gamma^{q}{}_{[tp]}a_{sq}+\Gamma^{q}_{s[t{}a_{p}]q}=0$,\\

where, the Riemannian curvature tensor is denoted by $R_{stpq}$, $e=\pm1$, and anti-symmetrization is indicated by square brackets. It is widely proposed in the scientific community that the Karmarkar condition can be used as a handy and simple way to compute the components $e^{\nu(r)}$ and $e^{\lambda(r)}$ of a spherically symmetric space-time. A prerequisite that is enough for the embedding of a class-I metric is given by this condition. Here is how the Karmarkar condition can be described:
\begin{equation}\label{15}
R_{0101}R_{2323}=R_{0202}R_{1313}-R_{1202}R_{1303}.
\end{equation}
Regarding metric (\ref{3}), the non-zero elements of the Riemannian tensor are as follows:
\begin{eqnarray*}
R_{0101}&=&-\frac{1}{4}e^{\nu(r)}\left(-a'(r)\lambda'(r)+\nu'^{2}(r)+2\nu''(r)\right),\quad
R_{2323}=-r^{2}\sin^{2}\theta\left(1-e^{-\lambda(r)}\right),\quad\\
R_{0202}&=&-\frac{1}{2}r\nu'(r)e^{\nu(r)-\lambda(r)},
R_{1313}=-\frac{1}{2}\lambda'(r) r\sin^{2}\theta,\quad
R_{1202}=0,\quad
R_{1303}=0.
\end{eqnarray*}
By just putting these Riemannian tensor constituent values into Eq. (\ref{15}), the differential equation that follows can be constructed with ease:
\begin{equation}\label{16}
\left(\lambda'(r)-\nu'(r)\right) \nu'(r)e^{\lambda(r)}+ 2\left(1-e^{\lambda(r)}\right)\nu''(r)+\nu'^2(r)=0.
\end{equation}
The fact that class-I embedding solutions can be obtained from Eq. (\ref{16}) and inserted into a 5-dimensional Euclidean space is an intriguing point to make. Eq. (\ref{16}) for $e^a$ can be integrated to obtain its expression in terms of $e^b$, as follows:
\begin{equation}\label{17}
e^{\nu(r)}=\left(A+B \int \sqrt{e^{\lambda(r)}-1} \, dr\right)^2,
\end{equation}
where A and B are constants for integration. The metric potentials $e^{\nu(r)}$ and $e^{\lambda(r)}$ are shown to be connected to one another from Eq. (\ref{17}). Consequently, this connection can be used to compute the other metric components if one of them is known. To build a realistic anisotropic model, we take into consideration a well-known model \cite{gedela2019new} for the $g_{rr}$ component of the metric, which is provided by:
\begin{eqnarray}\label{33aa}
\lambda(r)&=&\log \left(1+a r^2 e^{\kappa\sin ^{-1}\left(b r^2+c\right)}\right)
\end{eqnarray}
where $\kappa$, $b$, $c$, and $a$ notions the arbitrary constants. Reference \cite{gedela2019new} makes clear that, for a given range of parameters related to this $g_{rr}$ metric component, the solution given in Eq. (\ref{33aa}) is regular, correct, and yields a decent estimation of the neutron model. Eq. (\ref{33aa}) can be substituted into Eq. (\ref{17}) to yield:
\begin{eqnarray}\label{34aa}
\nu(r) &=& \log \left(\frac{B \left(\kappa  \sqrt{1-\left(b r^2+c\right)^2}+2 b r^2+2 c\right) \sqrt{a r^2 e^{\kappa  \sin ^{-1}\left(b r^2+c\right)}}}{b \left(\kappa ^2+4\right) r}+A\right)^2,
\end{eqnarray}
here, the values of the variables $A$, $B$, $\kappa$, $b$, $c$, and $a$ are selected for defining this space-time geometry according to various physical considerations. According to literature \cite{biswas2021anisotropic,zubair2021physical,rahaman2012strange,roupas2020anisotropic,maurya2023anisotropic}, this geometry has demonstrated efficacy in modeling self-gravitating stellar structures for both GR and modified gravity. \textcolor{black}{The inclusion of Eisenhart classes and the Karmarkar condition in this study is purposeful, as it constrains the spacetime geometry and facilitates consistent forms for the metric potentials \( \nu(r) \) and \( \lambda(r) \), which were confirmed through substitution into the \( f(T) \) field equations. The solutions in Eqs.~(\ref{33aa}) and (\ref{34aa}) exhibit insensitivity to the specific \( f(T) \) model as a result of the simplifying geometric constraints, yet they are consistent with \( f(T) \) gravity.} Employing Eqs. (\ref{33aa}-\ref{34aa}) along with Eqs. (\ref{28aa}-\ref{29aa}) into Eqs. (\ref{25aa}-\ref{27aa}), we yield:
\begin{eqnarray}
\rho+\rho^D &=&,\frac{(6 \gamma -1) }{4 \pi  (4 \gamma -1) r^2}\Big[-\frac{\beta  2^{n-2} n \left(1-\mathbb{Z}_8(r)\right) \left(\mathbb{Z}_4(r)\right){}^{n-1}}{a r^2 e^{\kappa  \sin ^{-1}\left(b r^2+c\right)}+1}+\frac{\beta  2^{n-2} n \left(\mathbb{Z}_3(r)\right) \left(\mathbb{Z}_4(r)\right){}^{n-1}}{a r^2 e^{\kappa  \sin ^{-1}\left(b r^2+c\right)}+1}+\beta  2^{n-2} r^2 \left(\mathbb{Z}_4(r)\right){}^n\nonumber\\&+&\beta  2^{n-2} (1-n) n r r \left(-\mathbb{Z}_9\right) \left(1-\mathbb{Z}_9(r)\right) \left(\mathbb{Z}_4(r)\right){}^{n-2}+\gamma  r^2 \psi  \log \left(2^{\chi } \phi  \left(\mathbb{Z}_4(r)\right){}^{\chi }\right)\Big],\label{35aa}\\
p+p_r^D &=& \frac{6 \gamma -1}{4 \pi  (4 \gamma -1)}\Big[-\frac{\beta  2^{n-1} n \left(\mathbb{Z}_4(r)\right){}^{n-1} \left(\sqrt{a r^2 e^{\kappa  \sin ^{-1}\left(b r^2+c\right)}+1}+\mathbb{Z}_5(r)-1\right)}{r^2 \left(a r^2 e^{\kappa  \sin ^{-1}\left(b r^2+c\right)}+1\right)}-\beta  2^{n-2} \left(\mathbb{Z}_4(r)\right){}^n\nonumber\\&-&\gamma  \psi  \log \left(2^{\chi } \phi  \left(\mathbb{Z}_4(r)\right){}^{\chi }\right)\Big],\label{36aa}\\
p+p_t^D &=& \frac{6 \gamma -1}{4 \pi  (4 \gamma -1)}\Big[\frac{\beta  2^{n-2} n \left(\mathbb{Z}_4(r)\right){}^{n-1}}{r^2 \left(a r^2 e^{\kappa  \sin ^{-1}\left(b r^2+c\right)}+1\right)} \Big[\frac{1}{4} r \Big[\frac{4 b B \left(\kappa ^2+4\right) \left(a r^2 e^{\kappa  \sin ^{-1}\left(b r^2+c\right)}\right)^{3/2}}{r \left(\mathbb{Z}_1(r)\right){}^2 \sqrt{1-\left(b r^2+c\right)^2}} \Big[B \Big[b^2 \kappa  r^4\nonumber\\&-&2 b r^2 \mathbb{Z}_{10}(r)-c^2 \kappa +2 c \mathbb{Z}_{10}(r)+\kappa \Big] \sqrt{a r^2 e^{\kappa  \sin ^{-1}\left(b r^2+c\right)}}+A b \left(\kappa ^2+4\right) r \left(b \kappa  r^2+r \mathbb{Z}_{10}\right)\Big]+\left(\mathbb{Z}_6(r)+2\right)\nonumber\\&\times& \Big[\frac{2 a b B \left(\kappa ^2+4\right) r^2 e^{\kappa  \sin ^{-1}\left(b r^2+c\right)}}{\mathbb{Z}_1(r)}-\mathbb{Z}_7(r)\Big]\Big]+\left(\mathbb{Z}_2(r)\right) \left(\sqrt{a r^2 e^{\kappa  \sin ^{-1}\left(b r^2+c\right)}+1}-1\right)\Big]\nonumber\\&-&\frac{\beta  2^{n-4} (1-n) n \left(\mathbb{Z}_4(r)\right){}^{n-2} \left(-2 \sqrt{a r^2 e^{\kappa  \sin ^{-1}\left(b r^2+c\right)}+1}+\mathbb{Z}_6(r)+2\right)}{r \left(a r^2 e^{\kappa  \sin ^{-1}\left(b r^2+c\right)}+1\right)}-\beta  2^{n-2} \left(\mathbb{Z}_4(r)\right){}^n\nonumber\\&-&\gamma  \psi  \log \left(2^{\chi } \phi  \left(\mathbb{Z}_4(r)\right){}^{\chi }\right)\Big]\label{37aa},
\end{eqnarray}
where,
\begin{eqnarray*}
    \mathbb{Z}_1(r)&=&A b \left(\kappa ^2+4\right) \sqrt{a r^2 e^{\kappa  \sin ^{-1}\left(b r^2+c\right)}}+a B r \left(\kappa  \sqrt{-b^2 r^4-2 b c r^2-c^2+1}+2 b r^2+2 c\right) e^{\kappa  \sin ^{-1}\left(b r^2+c\right)},\\
   \mathbb{Z}_2(r)&=&-\frac{2 a b B \left(\kappa ^2+4\right) r^3 e^{\kappa  \sin ^{-1}\left(b r^2+c\right)}}{\mathbb{Z}_1(r)}+\sqrt{a r^2 e^{\kappa  \sin ^{-1}\left(b r^2+c\right)}+1}-1,\\
   \mathbb{Z}_3(r)&=&\frac{2 a b B \left(\kappa ^2+4\right) r^3 e^{\kappa  \sin ^{-1}\left(b r^2+c\right)} \left(\sqrt{a r^2 e^{\kappa  \sin ^{-1}\left(b r^2+c\right)}+1}-1\right)}{\mathbb{Z}_1(r)}+2 \sqrt{a r^2 e^{\kappa  \sin ^{-1}\left(b r^2+c\right)}+1}-1,\\
  \mathbb{Z}_4(r)&=&\frac{\left(\mathbb{Z}_2(r)\right) \left(\sqrt{a r^2 e^{\kappa  \sin ^{-1}\left(b r^2+c\right)}+1}-1\right)}{r^2 \left(a r^2 e^{\kappa  \sin ^{-1}\left(b r^2+c\right)}+1\right)},\; \mathbb{Z}_6(r)=\frac{2 a b B \left(\kappa ^2+4\right) r^3 e^{\kappa  \sin ^{-1}\left(b r^2+c\right)}}{\mathbb{Z}_1(r)},\\
  \mathbb{Z}_5(r)&=&\frac{a b B \left(\kappa ^2+4\right) r^3 e^{\kappa  \sin ^{-1}\left(b r^2+c\right)} \left(\sqrt{a r^2 e^{\kappa  \sin ^{-1}\left(b r^2+c\right)}+1}-2\right)}{\mathbb{Z}_1(r)},
\end{eqnarray*}
\begin{eqnarray*}
  \mathbb{Z}_7(r)&=&\frac{2 a r e^{\kappa  \sin ^{-1}\left(b r^2+c\right)}+\frac{2 a b \kappa  r^3 e^{\kappa  \sin ^{-1}\left(b r^2+c\right)}}{\sqrt{1-\left(b r^2+c\right)^2}}}{a r^2 e^{\kappa  \sin ^{-1}\left(b r^2+c\right)}+1},\;\mathbb{Z}_8(r)=\frac{2 a r^2 \left(\frac{b \kappa  r^2}{\sqrt{1-\left(b r^2+c\right)^2}}+1\right) e^{\kappa  \sin ^{-1}\left(b r^2+c\right)}}{a r^2 e^{\kappa  \sin ^{-1}\left(b r^2+c\right)}+1},\\
  \mathbb{Z}_9(r)&=&\frac{1}{\sqrt{a r^2 e^{\kappa  \sin ^{-1}\left(b r^2+c\right)}+1}},\;\mathbb{Z}_{10}(r)=\sqrt{-b^2 r^4-2 b c r^2-c^2+1}.
\end{eqnarray*}
\textcolor{black}{The equations for density and pressure presented in Eqs.~(\ref{35aa}-\ref{37aa}) establish a general framework for stellar configurations without specifying a particular equation of state (EoS). It is recognized that for the modeling of white dwarfs, the inclusion of the degenerate Fermi gas equation of state, as referenced in \cite{de2023mass}, is essential, and observationally motivated equations of state will considered in future extensions. We are currently concentrating on developing a general theoretical framework that can be adapted to different stellar scenarios.} Eqs (\ref{35aa}-\ref{37aa}) can be solved to yield the expressions of pressures and energy density pertaining the dark energy matter and ordinary matter separately. Assume for the moment that the dark energy density, $\left(\rho^D\right)$, is proportional to the radial pressure associated with it:
$$
p_r^D=-\rho^D
$$
Where the typical baryonic matter density is proportionate to the energy density associated with dark energy, that is,
$$
\rho^D=\omega \rho
$$
The earlier research work listed in references \cite{ghezzi2011anisotropic,bhar2023dark,das2023acceptable} served as inspiration for this strategy. In this case, $\omega$ is a non-zero constant that is taken as a free parameter. The energy density and pressure for typical baryonic matter are now derived by solving the aforementioned set of equations that are listed below:

\begin{eqnarray}
    \rho &=&\frac{6 \gamma -1}{4 \pi  (4 \gamma -1) r^2 (\omega +1)}\Big[-\frac{\beta  2^{n-2} n \left(1-\mathbb{Z}_8(r)\right) \left(\mathbb{Z}_4(r)\right){}^{n-1}}{a r^2 e^{\kappa  \sin ^{-1}\left(b r^2+c\right)}+1}+\frac{\beta  2^{n-2} n \left(\mathbb{Z}_3(r)\right) \left(\mathbb{Z}_4(r)\right){}^{n-1}}{a r^2 e^{\kappa  \sin ^{-1}\left(b r^2+c\right)}+1}\nonumber\\&+&\beta  2^{n-2} r^2 \left(\mathbb{Z}_4(r)\right){}^n+\beta  2^{n-2} (1-n) n r  \left(-\mathbb{Z}_9(r)\right) \left(1-\mathbb{Z}_9(r)\right) \left(\mathbb{Z}_4(r)\right){}^{n-2}+\gamma  r^2 \psi  \log \left(2^{\chi } \phi  \left(\mathbb{Z}_4(r)\right){}^{\chi }\right)\Big],\\
    p &=&\frac{(6 \gamma -1) \omega }{4 \pi  (4 \gamma -1) r^2 (\omega +1)} \Big[-\frac{\beta  2^{n-2} n \left(1-\mathbb{Z}_8(r)\right) \left(\mathbb{Z}_4(r)\right){}^{n-1}}{a r^2 e^{\kappa  \sin ^{-1}\left(b r^2+c\right)}+1}+\frac{\beta  2^{n-2} n \left(\mathbb{Z}_3(r)\right) \left(\mathbb{Z}_4(r)\right){}^{n-1}}{a r^2 e^{\kappa  \sin ^{-1}\left(b r^2+c\right)}+1}\nonumber\\&+&\beta  2^{n-2} r^2 \left(\mathbb{Z}_4(r)\right){}^n+\beta  2^{n-2} (1-n) n r \left(-\mathbb{Z}_9(r)\right) \left(1-\mathbb{Z}_9(r)\right) \left(\mathbb{Z}_4(r)\right){}^{n-2}+\gamma  r^2 \psi  \log \left(2^{\chi } \phi  \left(\mathbb{Z}_4(r)\right){}^{\chi }\right)\Big]\nonumber\\&+&\frac{(6 \gamma -1)}{4 \pi  (4 \gamma -1)} \Big[-\frac{\beta  2^{n-1} n \left(\mathbb{Z}_4(r)\right){}^{n-1} \left(\sqrt{a r^2 e^{\kappa  \sin ^{-1}\left(b r^2+c\right)}+1}+\mathbb{Z}_5(r)-1\right)}{r^2 \left(a r^2 e^{\kappa  \sin ^{-1}\left(b r^2+c\right)}+1\right)}-\beta  2^{n-2} \left(\mathbb{Z}_4(r)\right){}^n\nonumber\\&-&\gamma  \psi  \log \left(2^{\chi } \phi  \left(\mathbb{Z}_4(r)\right){}^{\chi }\right)\Big],\\
    \rho^D &=& \frac{6 \gamma -1}{4 \pi  (4 \gamma -1) r^2}\Big[-\frac{\beta  2^{n-2} n \left(1-\mathbb{Z}_8(r)\right) \left(\mathbb{Z}_4(r)\right){}^{n-1}}{a r^2 e^{\kappa  \sin ^{-1}\left(b r^2+c\right)}+1}+\frac{\beta  2^{n-2} n \left(\mathbb{Z}_3(r)\right) \left(\mathbb{Z}_4(r)\right){}^{n-1}}{a r^2 e^{\kappa  \sin ^{-1}\left(b r^2+c\right)}+1}+\beta  2^{n-2} r^2 \left(\mathbb{Z}_4(r)\right){}^n\nonumber\\&+&\beta  2^{n-2} (1-n) n r \left(-\mathbb{Z}_9(r)\right) \left(1-\mathbb{Z}_9(r)\right) \left(\mathbb{Z}_4(r)\right)^{n-2}+\gamma  r^2 \psi  \log \left(2^{\chi } \phi  \left(\mathbb{Z}_4(r)\right)^{\chi }\right)\Big]-\frac{6 \gamma -1}{4 \pi  (4 \gamma -1) r^2 (\omega +1)} \nonumber\\&\times& \Big[-\frac{\beta  2^{n-2} n \left(1-\mathbb{Z}_8(r)\right) \left(\mathbb{Z}_4(r)\right){}^{n-1}}{a r^2 e^{\kappa  \sin ^{-1}\left(b r^2+c\right)}+1}+\frac{\beta  2^{n-2} n \left(\mathbb{Z}_3(r)\right) \left(\mathbb{Z}_4(r)\right){}^{n-1}}{a r^2 e^{\kappa  \sin ^{-1}\left(b r^2+c\right)}+1}+\beta  2^{n-2} r^2 \left(\mathbb{Z}_4(r)\right)^n \nonumber\\&+&\beta  2^{n-2} (1-n) n r r \left(-\mathbb{Z}_9\right) \left(1-\mathbb{Z}_9(r)\right) \left(\mathbb{Z}_4(r)\right){}^{n-2}+\gamma  r^2 \psi  \log \left(2^{\chi } \phi  \left(\mathbb{Z}_4(r)\right){}^{\chi }\right)\Big],\\
    p_r^D&=&-\frac{(6 \gamma -1) \omega }{4 \pi  (4 \gamma -1) r^2 (\omega +1)}\Big[-\frac{\beta  2^{n-2} n \left(1-\mathbb{Z}_8(r)\right) \left(\mathbb{Z}_4(r)\right){}^{n-1}}{a r^2 e^{\kappa  \sin ^{-1}\left(b r^2+c\right)}+1}+\frac{\beta  2^{n-2} n \left(\mathbb{Z}_3(r)\right) \left(\mathbb{Z}_4(r)\right){}^{n-1}}{a r^2 e^{\kappa  \sin ^{-1}\left(b r^2+c\right)}+1}\nonumber\\&+&\beta  2^{n-2} r^2 \left(\mathbb{Z}_4(r)\right){}^n+\beta  2^{n-2} (1-n) n r \left(-\mathbb{Z}_9(r)\right) \left(1-\mathbb{Z}_9(r)\right) \left(\mathbb{Z}_4(r)\right){}^{n-2}+\gamma  r^2 \psi  \log \left(2^{\chi } \phi  \left(\mathbb{Z}_4(r)\right){}^{\chi }\right)\Big],
\end{eqnarray}
\begin{eqnarray}
    p_t^D&=&\frac{(6 \gamma -1)}{4 \pi  (4 \gamma -1)} \Big[\frac{\beta  2^{n-2} n \left(\mathbb{Z}_4(r)\right){}^{n-1}}{r^2 \left(a r^2 e^{\kappa  \sin ^{-1}\left(b r^2+c\right)}+1\right)} \Big[\frac{1}{4} r \Big[\frac{4 b B \left(\kappa ^2+4\right) \left(a r^2 e^{\kappa  \sin ^{-1}\left(b r^2+c\right)}\right)^{3/2}}{r \left(\mathbb{Z}_1(r)\right){}^2 \sqrt{1-\left(b r^2+c\right)^2}} \Big[B \Big[b^2 \kappa  r^4\nonumber\\&-&2 b r^2 \mathbb{Z}_{10}(r)-c^2 \kappa +2 c \mathbb{Z}_{10}(r)+\kappa \Big] \sqrt{a r^2 e^{\kappa  \sin ^{-1}\left(b r^2+c\right)}}+A b \left(\kappa ^2+4\right) r \left(b \kappa  r^2+ \right)\Big]\nonumber\\&+&\left(\mathbb{Z}_6(r)+2\right) \Big[\frac{2 a b B \left(\kappa ^2+4\right) r^2 e^{\kappa  \sin ^{-1}\left(b r^2+c\right)}}{\mathbb{Z}_1(r)}-\mathbb{Z}_7(r)\Big]\Big]+\left(\mathbb{Z}_2(r)\right) \left(\sqrt{a r^2 e^{\kappa  \sin ^{-1}\left(b r^2+c\right)}+1}-1\right)\Big]\nonumber\\&-&\frac{\beta  2^{n-4} (1-n) n \left(\mathbb{Z}_4(r)\right){}^{n-2} \left(-2 \sqrt{a r^2 e^{\kappa  \sin ^{-1}\left(b r^2+c\right)}+1}+\mathbb{Z}_6(r)+2\right)}{r \left(a r^2 e^{\kappa  \sin ^{-1}\left(b r^2+c\right)}+1\right)}-\beta  2^{n-2} \left(\mathbb{Z}_4(r)\right){}^n\nonumber\\&-&\gamma  \psi  \log \left(2^{\chi } \phi  \left(\mathbb{Z}_4(r)\right){}^{\chi }\right)\Big]-\frac{(6 \gamma -1) \omega }{4 \pi  (4 \gamma -1) r^2 (\omega +1)} \Big[-\frac{\beta  2^{n-2} n \left(1-\mathbb{Z}_8(r)\right) \left(\mathbb{Z}_4(r)\right){}^{n-1}}{a r^2 e^{\kappa  \sin ^{-1}\left(b r^2+c\right)}+1}\nonumber\\&+&\frac{\beta  2^{n-2} n \left(\mathbb{Z}_3(r)\right) \left(\mathbb{Z}_4(r)\right){}^{n-1}}{a r^2 e^{\kappa  \sin ^{-1}\left(b r^2+c\right)}+1}+\beta  2^{n-2} r^2 \left(\mathbb{Z}_4(r)\right){}^n+\beta  2^{n-2} (1-n) n  r \left(-\mathbb{Z}_9(r)\right) \left(1-\mathbb{Z}_9(r)\right) \left(\mathbb{Z}_4(r)\right){}^{n-2}\nonumber\\&+&\gamma  r^2 \psi  \log \left(2^{\chi } \phi  \left(\mathbb{Z}_4(r)\right){}^{\chi }\right)\Big]-\frac{(6 \gamma -1) }{4 \pi  (4 \gamma -1)}\Big[-\frac{\beta  2^{n-1} n \left(\mathbb{Z}_4(r)\right){}^{n-1} \left(\sqrt{a r^2 e^{\kappa  \sin ^{-1}\left(b r^2+c\right)}+1}+\mathbb{Z}_5(r)-1\right)}{r^2 \left(a r^2 e^{\kappa  \sin ^{-1}\left(b r^2+c\right)}+1\right)}\nonumber\\&-&\beta  2^{n-2} \left(\mathbb{Z}_4(r)\right){}^n-\gamma  \psi  \log \left(2^{\chi } \phi  \left(\mathbb{Z}_4(r)\right){}^{\chi }\right)\Big].
\end{eqnarray}

In our study, we find solutions for pressure and energy density using an EoS customized to dark energy stars, which takes dark energy into primary consideration. This EoS originates from a different source than the standard neutron star EoS, which is based on nuclear and quantum chromodynamic interactions. We instead use theoretical dark energy models in our EoS, which usually have negative pressure and constant energy density. This EoS does not arise directly from physics of fundamental particles, but it provides a sound theoretical basis for investigating dark energy star characteristics. While our knowledge of dark energy develops, we recognize that our model is theoretical and that it may be improved by comparing it with more precise particle physics models and empirical facts.

\begin{figure}
\includegraphics[width=5.7cm, height=4.5cm]{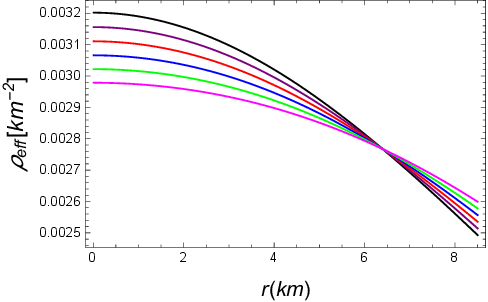}
\includegraphics[width=5.7cm, height=4.5cm]{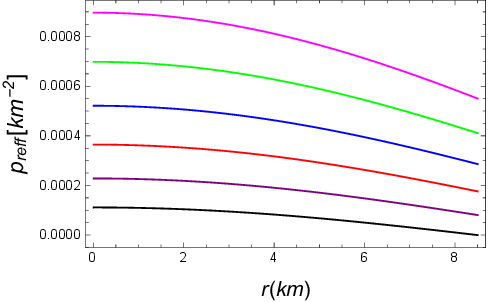}
\includegraphics[width=5.7cm, height=4.5cm]{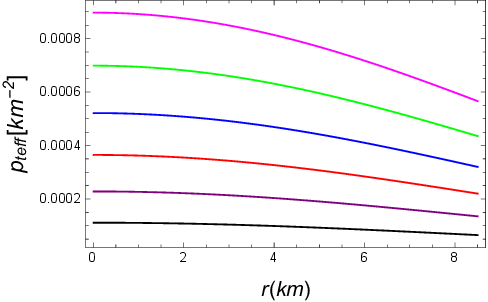}
\caption{The graphical profiles of $\rho_{eff},\;p_{reff},\;p_{teff}$ evolving along the radial coordinate $r$ for $\kappa=1.0(\text{black}),\;\kappa=1.2(\text{purple}),\;\kappa=1.4(\text{red}),\;\kappa=1.6(\text{blue}),\;\kappa=1.0(\text{green})\;\&\;\kappa=1.0(\text{magenta})$ for compact star candidate $Her\;X-1$. \textcolor{black}{Other fixed parameters  $n,\;\beta,\;\gamma,\;\phi,\;\psi,\;b,\; c ,\;\chi\;\&\;\omega $ are mentioned in Table \ref{tab1}}.}
\end{figure}\label{fig:1}

\begin{figure}
\includegraphics[width=4.0cm, height=3.8cm]{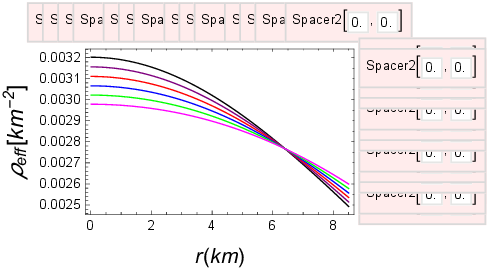}
\includegraphics[width=1.7cm, height=3.8cm]{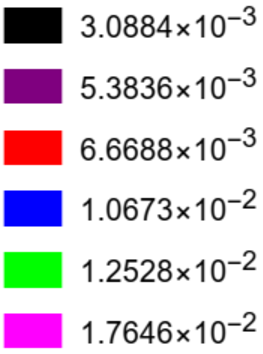}
\includegraphics[width=4.0cm, height=3.8cm]{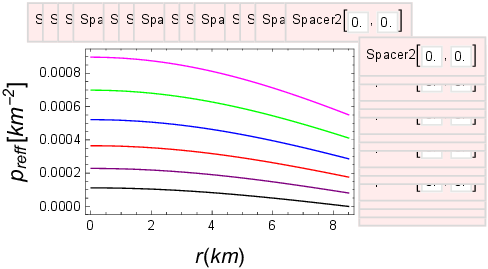}
\includegraphics[width=1.7cm, height=3.8cm]{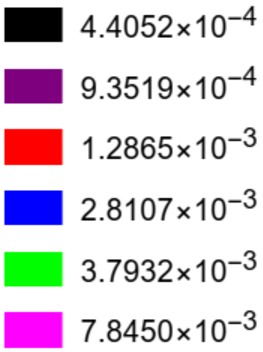}
\includegraphics[width=4.0cm, height=3.8cm]{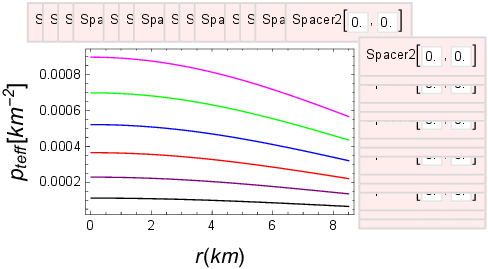}
\includegraphics[width=1.7cm, height=3.8cm]{FigNa2.eps}
\caption{The graphical profiles of central density $\rho_{effc}$, central pressures $p_{reffc}\;\&\;p_{teffc}$ evolving along radial coordinate $r$ for compact stars candidates. $Her\;X-1(\text{black}),\;LMC\;X-4(\text{purple}),\;Cen\;X-3(\text{Red}),\;PSR\;J1614-2230(\text{blue}),\;PSR\;J0740+6620(\text{green})
\;\&\;GW190814(\text{black})magenta$.}
\end{figure}\label{fig:2}

\begin{figure}
\centering
\includegraphics[width=7cm, height=5cm]{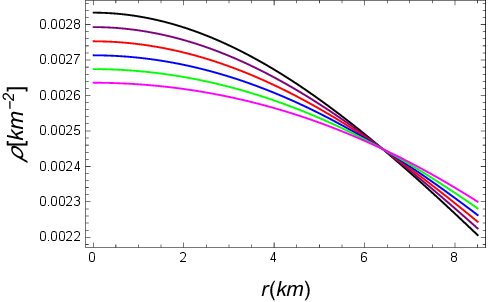}
\includegraphics[width=7cm, height=5cm]{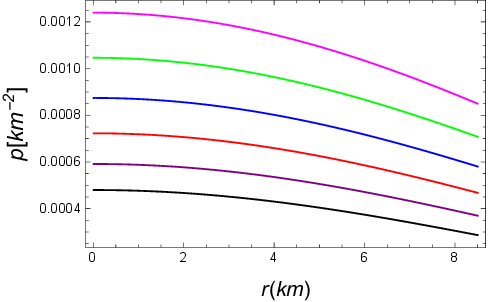}
\caption{Graph showing the profiles of density component $\rho$ and pressure component $p$ evolving versus the radial coordinate $r$ for $\kappa=1.0(\text{black}),\;\kappa=1.2(\text{purple}),\;\kappa=1.4(\text{red}),\;\kappa=1.6(\text{blue}),\;\kappa=1.0(\text{green})\;\&\;\kappa=1.0(\text{magenta})$ for the celestial candidate $Her\;X-1$. \textcolor{black}{Other fixed parameters  $n,\;\beta,\;\gamma,\;\phi,\;\psi,\;b,\; c ,\;\chi\;\&\;\omega $ are mentioned in Table \ref{tab1}}.}
\end{figure}\label{fig:3}

\begin{figure}
\centering
\includegraphics[width=5.7cm, height=4.5cm]{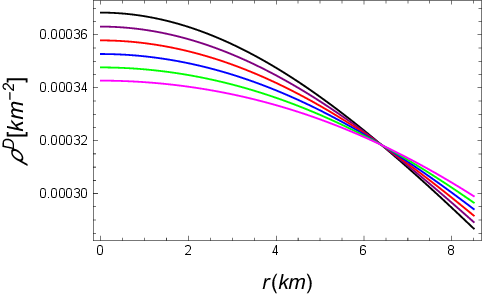}
\includegraphics[width=5.7cm, height=4.5cm]{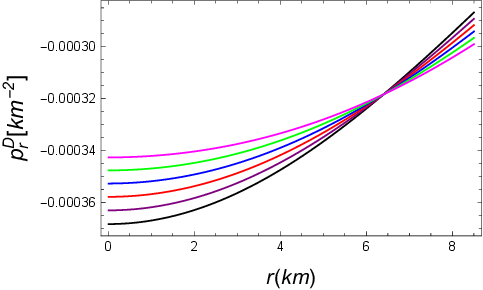}
\includegraphics[width=5.7cm, height=4.5cm]{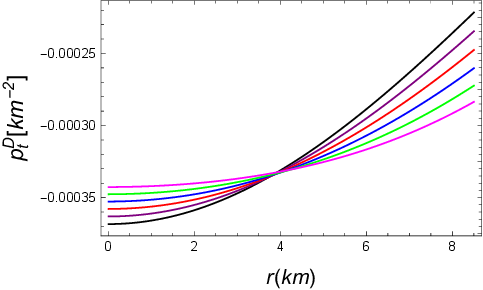}
\caption{The graphical profiles of dark energy density component $\rho^{D}$, dark energy radial pressure component $p_r^D$, and dark energy tangential pressure $p_t^D$ evolving versus the radial coordinate $r$ for $\kappa=1.0(\text{black}),\;\kappa=1.2(\text{purple}),\;\kappa=1.4(\text{red}),\;\kappa=1.6(\text{blue}),\;\kappa=1.0(\text{green})\;\&\;\kappa=1.0(\text{magenta})$ for compact star candidate $Her\;X-1$. \textcolor{black}{Other fixed parameters  $n,\;\beta,\;\gamma,\;\phi,\;\psi,\;b,\; c ,\;\chi\;\&\;\omega $ are mentioned in Table \ref{tab1}}. }
\end{figure}\label{fig:4}
\begin{figure}
\centering
\includegraphics[width=5.7cm, height=4.5cm]{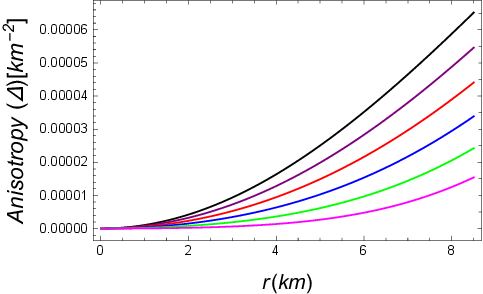}
\includegraphics[width=5.7cm, height=4.5cm]{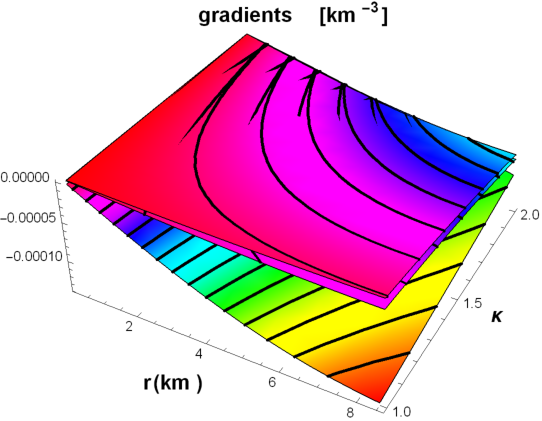}
\includegraphics[width=5.0cm, height=4.5cm]{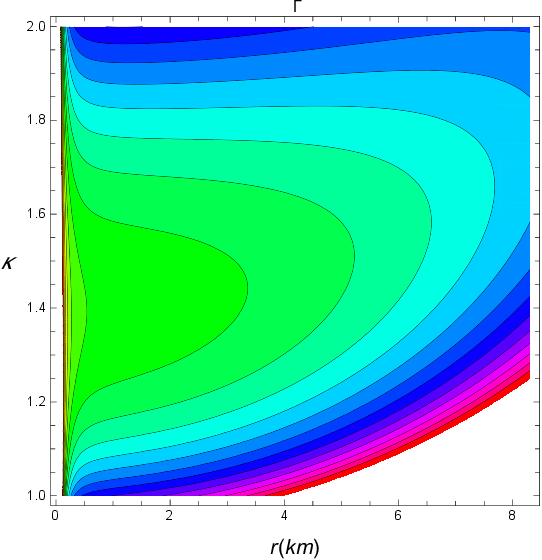}
\includegraphics[width=0.7cm, height=4.4cm]{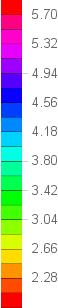}
\caption{The graphical profiles of anisotropy left graph evolving along radial coordinate $r$ for $\kappa=1.0(\text{black}),\;\kappa=1.2(\text{purple}),\;\kappa=1.4(\text{red}),\;\kappa=1.6(\text{blue}),\;\kappa=1.0(\text{green})\;\&\;\kappa=1.0(\text{magenta})$ for celestial candidate $Her\;X-1$. Gradients and adiabatic index evolve in the middle and right graph, respectively, versus the radial coordinate $r$ for a range of $\kappa=1-2$. \textcolor{black}{Other fixed parameters  $n,\;\beta,\;\gamma,\;\phi,\;\psi,\;b,\; c ,\;\chi\;\&\;\omega $ are mentioned in Table \ref{tab1}}.}
\end{figure}\label{fig:5}

\begin{figure}
\centering
\includegraphics[width=5.7cm, height=4.5cm]{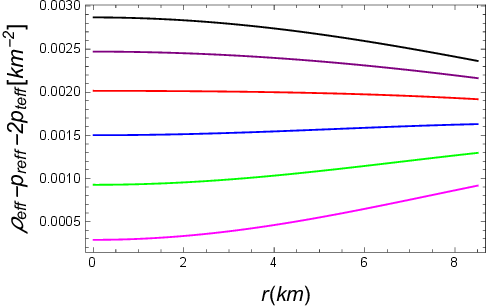}
\includegraphics[width=5.7cm, height=4.5cm]{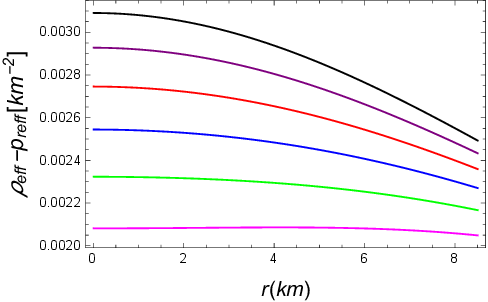}
\includegraphics[width=5.7cm, height=4.5cm]{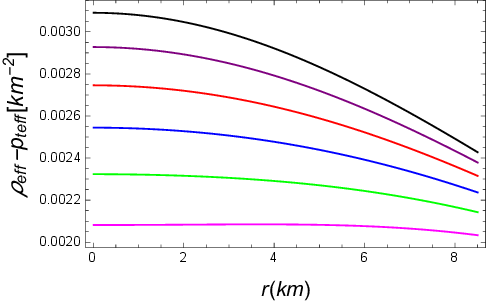}\\
\includegraphics[width=5.7cm, height=4.5cm]{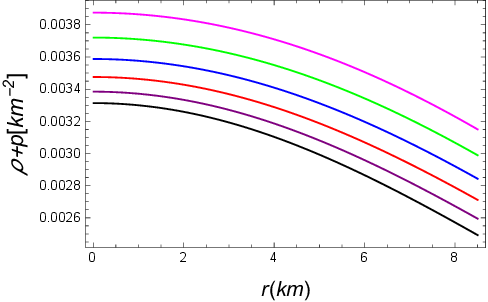}
\includegraphics[width=5.7cm, height=4.5cm]{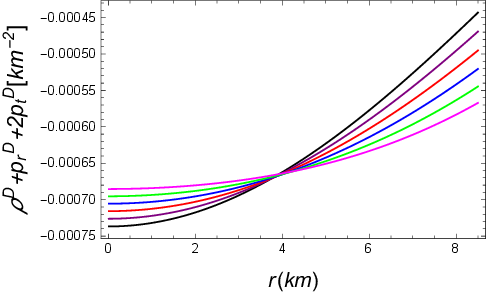}
\caption{The graphical profiles of energy conditions evolving along the radial coordinate $r$ for $\kappa=1.0(\text{black}),\;\kappa=1.2(\text{purple}),\;\kappa=1.4(\text{red}),\;\kappa=1.6(\text{blue}),\;\kappa=1.0(\text{green})\;\&\;\kappa=1.0(\text{magenta})$ for compact star candidate $Her\;X-1$. \textcolor{black}{Other fixed parameters  $n,\;\beta,\;\gamma,\;\phi,\;\psi,\;b,\; c ,\;\chi\;\&\;\omega $ are mentioned in Table \ref{tab1}}.}
\end{figure}\label{fig:6}

\begin{figure}
\centering
\includegraphics[width=6.7cm, height=5.0cm]{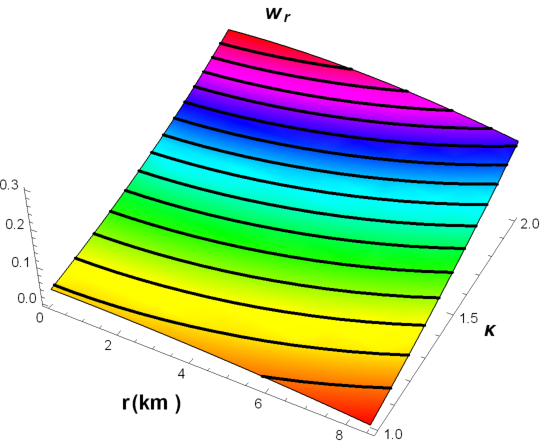}
\includegraphics[width=6.7cm, height=5.0cm]{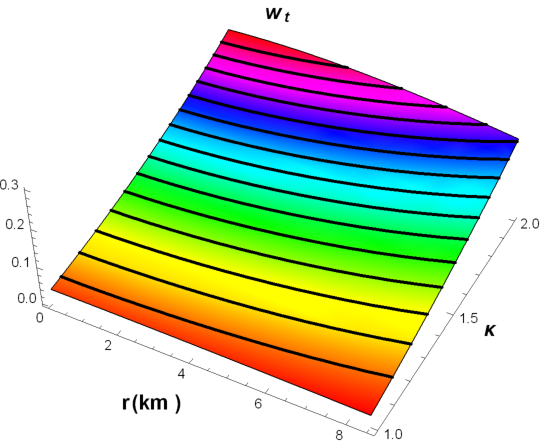}\\
\includegraphics[width=6.7cm, height=5.0cm]{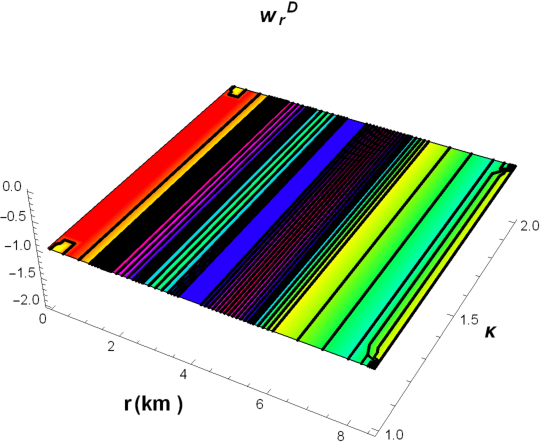}
\includegraphics[width=6.7cm, height=5.0cm]{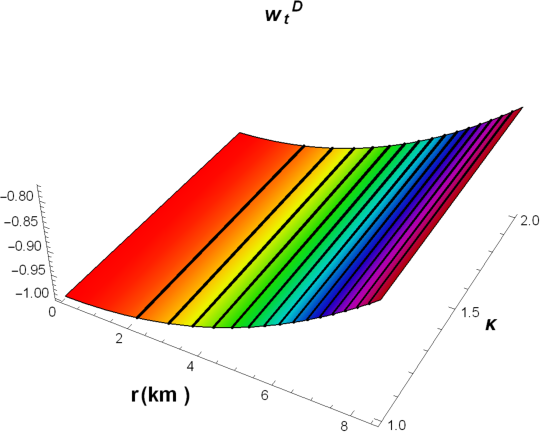}
\caption{Equation of state components for normal matter upper panel graphs, and dark matter lower two graphs along radial coordinate $r$ for a range of $\kappa=1-2$ for celestial candidate $Her\;X-1$. \textcolor{black}{Other fixed parameters  $n,\;\beta,\;\gamma,\;\phi,\;\psi,\;b,\; c ,\;\chi\;\&\;\omega $ are mentioned in Table \ref{tab1}}.}
\end{figure}\label{fig:7}

\begin{figure}
\centering
\includegraphics[width=5.7cm, height=4.5cm]{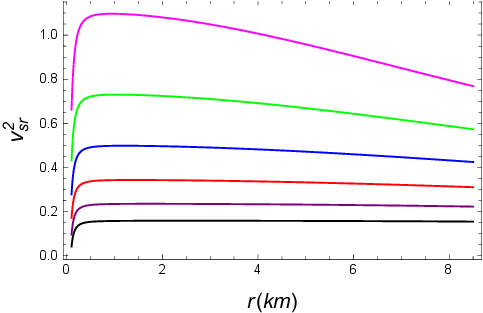}
\includegraphics[width=5.7cm, height=4.5cm]{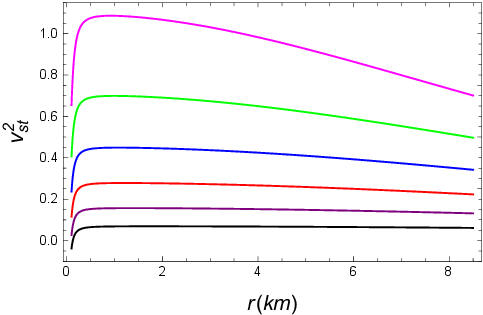}
\includegraphics[width=5.7cm, height=4.5cm]{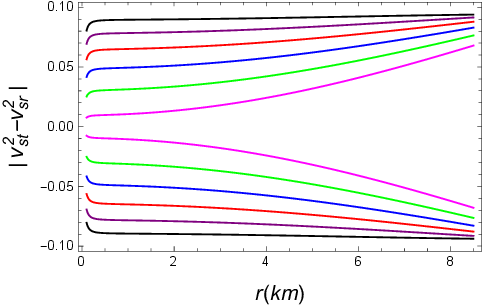}
\caption{The graphical profiles of sound speeds and Abreu limit evolving along the radial coordinate $r$ for $\kappa=1.0(\text{black}),\;\kappa=1.2(\text{purple}),\;\kappa=1.4(\text{red}),\;\kappa=1.6(\text{blue}),\;\kappa=1.0(\text{green})\;\&\;\kappa=1.0(\text{magenta})$ for compact star candidate $Her\;X-1$.  \textcolor{black}{Other fixed parameters  $n,\;\beta,\;\gamma,\;\phi,\;\psi,\;b,\; c ,\;\chi\;\&\;\omega $ are mentioned in Table \ref{tab1}}.}
\end{figure}\label{fig:8}

\begin{figure}
\centering
\includegraphics[width=5.7cm, height=4.5cm]{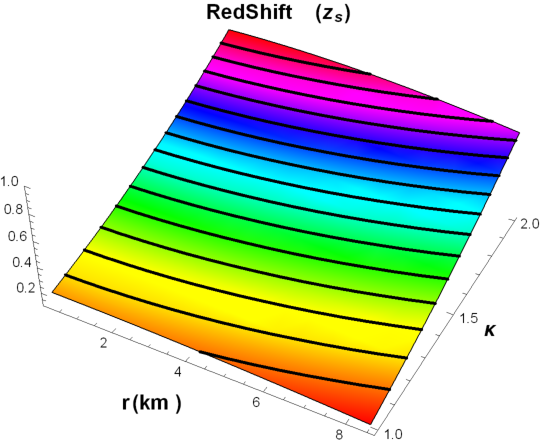}
\includegraphics[width=5.7cm, height=4.5cm]{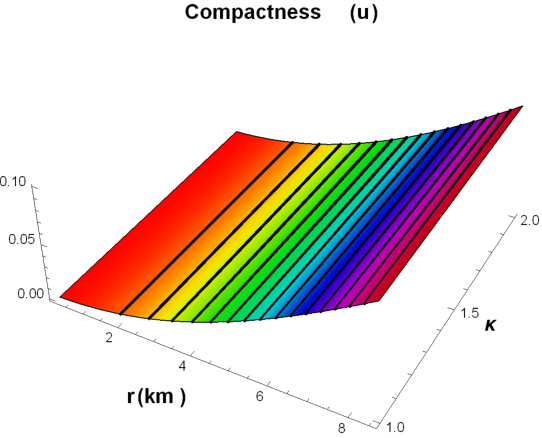}
\includegraphics[width=5.7cm, height=4.5cm]{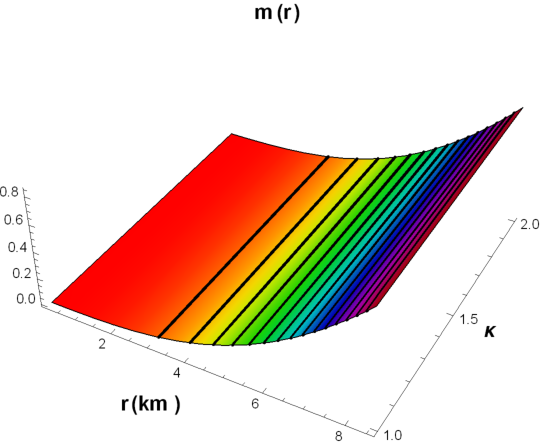}
\caption{The graphical profiles of redshift parameter,  compactness parameter, and mass function evolving along the radial coordinate $r$ for a range of $\kappa=1-2$ for compact star candidate $Her\;X-1$. \textcolor{black}{Other fixed parameters  $n,\;\beta,\;\gamma,\;\phi,\;\psi,\;b,\; c ,\;\chi\;\&\;\omega $ are mentioned in Table \ref{tab1}}.}
\end{figure}\label{fig:9}

\section{Matching Conditions}\label{sec3}
\textcolor{black}{This study investigates the alteration of the Schwarzschild-de Sitter metric in the framework of $f(T)$ gravity. Although we acknowledge the need to obtain an external solution that satisfies the boundary requirements, the system's complexity has constrained our capacity to fully accomplish this in the current research. Specifically, we observe that the exterior solution in weak-field regimes is accessible solely through perturbative methods for a power-law ansatz, as elaborated in \cite{bahamonde2023teleparallel}. Furthermore, we have examined a Rastall modification employing an ansatz of the type $h(T) = \psi \log(\varphi T^\chi)$, which complicates the matching process further.} An essential requirement while studying the compact celestial models is evaluating constant parameters involved in the study. In the present article, these unknowns are assessed by the uniform matching of inner space-time Eq. (\ref{1}) and outer vacuum spacetime, While discussion the $f(T)$ gravity framework the exterior Schwarzschild-de Sitter solution may be the best suited exterior spacetime:
{\color{black}
\begin{eqnarray}
\mathrm{d} s^2=-\left(1-\frac{2 M}{R}-\frac{\Lambda}{3}R^2\right) \mathrm{d} t^2+\left(1-\frac{2 M}{R}-\frac{\Lambda}{3}R^2\right)^{-1} \mathrm{~d} r^2-r^2\left(\mathrm{~d} \theta^2+\sin ^2 \theta \mathrm{d} \phi^2\right) .
\end{eqnarray}
Here M represents the total mass of the stellar object at the boundary. The observational value of $\Lambda$ is so small that at the boundary when $r=R$ may be neglected.} The following renowned Israel-Darmois junction criteria \cite{israel1966singular,darmois1927memorial} must be met for uniform and consistent matching:
\begin{eqnarray}
\left(\mathrm{d} s^2\right)^{+}=\left(\mathrm{d} s^2\right)^{-},\left(\kappa_{i j}\right)^{+}=\left(\kappa_{i j}\right)^{-}, \\
\left.e^{2 \lambda^{+}}\right|_{r=R}=\left.e^{2 \lambda^{+}}\right|_{r=R},\left.e^{2 v^{+}}\right|_{r=R}=\left.e^{2 v^{+}}\right|_{r=R}, \\
\left.\frac{\partial e^{2 v^{+}}}{\partial r}\right|_{r=R}=\left.\frac{\partial e^{2 v^{+}}}{\partial r}\right|_{r=R},\left.\& p_r\right|_{r=R}=0 .
\end{eqnarray}
Thus, if we consider the circumstances above, we obtain the subsequent set of equations:
\begin{eqnarray}
  \Big[A+\frac{B \left(n \sqrt{1-\left(b R^2+c\right)^2}+2 b R^2+2 c\right) \sqrt{a R^2 e^{n \sin ^{-1}\left(b R^2+c\right)}}}{b \left(n^2+4\right) R}\Big]^2&=&1-\frac{2 M}{R}-\frac{\Lambda}{3}R^2,\label{36}\\
    a R^2 e^{n \sin ^{-1}\left(b R^2+c\right)}+1&=&\frac{1}{1-\frac{2 M}{R}-\frac{\Lambda}{3}R^2},\label{37}\\
   2 a B R e^{n \sin ^{-1}\left(b R^2+c\right)} \Big[\frac{A R}{\sqrt{a R^2 e^{n \sin ^{-1}\left(b R^2+c\right)}}}+\frac{B \left(2 b R^2+2 c+n  \mathbb{Z}(R)_{10}\right)}{b \left(n^2+4\right)}\Big]&=&\frac{2 M}{R^2}-\frac{2\Lambda}{3}R.\label{38}
\end{eqnarray}
The subsequent formulations for the constants $a,\;A,\;B$ are obtained from the solution of above equations:
\begin{eqnarray}
    a&=&-\frac{2 M e^{-n \sin ^{-1}\left(b R^2+c\right)}}{R^2 (2 M-R)},\label{39}\\
    A&=&\frac{1}{b \left(n^2+4\right) R}\Big[b R \Big[-2 B R \sqrt{a R^2 e^{n \sin ^{-1}\left(b R^2+c\right)}}+n^2 \sqrt{1-\frac{2 M}{R}}+4 \sqrt{1-\frac{2 M}{R}}\Big]\nonumber\\&-&B \left(n \sqrt{-\left(b R^2+c-1\right) \left(b R^2+c+1\right)}+2 c\right) \sqrt{a R^2 e^{n \sin ^{-1}\left(b R^2+c\right)}}\Big],\label{40}\\
    B&=&\frac{M}{R^2 \sqrt{2-\frac{4 M}{R}} \sqrt{-\frac{M}{2 M-R}}}.\label{41}
\end{eqnarray}
The constant parameters' specific values are provided in Table-\ref{tab1}.
\begin{table*}
\caption{\label{tab1}{Values of constants compact star candidate $Her\;X-1\;(M=0.85\;\&\;R=8.5)$, by fixing $n=1,\;\beta=5,\;\gamma =1,\;\textcolor{black}{\phi =2},\;\psi =-2.0\times 10^{-5},\;b=0.001,\; c = 0.001,\;\chi =5\;\&\;\omega =0.13$.}}
\begin{center}
\begin{tabular}{ccccccccccccccc}
\hline
$\kappa$&  a & A &    B& $\frac{p_{reffc}}{\rho_{effc}}$ ($r=0$) &$\rho_{effc}$ &$p_{reffc}$ &$p_{teffc}$ &$\rho_{c}$&$p_{c}$&$\rho^{D}_{c}$ &$p_{rc}^{D}$ &$p_{tc}^{D}$\\
\hline
1.0 & 0.00321& 0.540428 & 0.0263                   &  $<1$ &0.00320 &0.000111 &0.000111&0.00283 &0.000479&0.000368&-0.000368 &-0.000368\\
\hline
 1.2& 0.003168&  0.439978  & 0.0263                 & $<1$ &0.00315  &0.000228 &0.000228& 0.00279&0.000591&0.000363& -0.000363&-0.000363\\
\hline
 1.4& 0.003122 & 0.349802  &   0.0263             &$<1$ & 0.00311 & 0.000364&0.000364& 0.00275&0.000722&0.000357& -0.000357&-0.000357\\
\hline
 1.6& 0.003077  & 0.270756   &   0.0263             &$<1$ &  0.00306& 0.000521&0.000521& 0.00271&0.000874&0.000352& -0.000352&-0.000352\\
\hline
 1.8& 0.003032 & 0.202694   &   0.0263             &$<1$ & 0.00302 & 0.000698&0.000698& 0.00267&0.001046&0.000347& -0.000347&-0.000347\\
\hline
 2.0& 0.002988 & 0.144856   &   0.02637             &$<1$ &  0.00297& 0.000896&0.000896& 0.00263&0.001239&0.000342& -0.000342&-0.000342\\
\hline
\end{tabular}
\end{center}
\end{table*}

\begin{table*}
\caption{\label{tab2}{Comparison of densities and pressures of different compact stars candidate, by fixing $\kappa=1.5$ $n=1,\;\beta=5,\;\gamma =1,\;\textcolor{black}{\phi =2},\;\psi =-2.0\times 10^{-5},\;b=0.001,\; c = 0.001,\;\chi =5\;\&\;\omega =0.13$.}}
\begin{center}
\begin{tabular}{ccccccccccccccc}
\hline
Compact star&  $M_{O}$ &$R\;(km)$ & $\frac{p_{reffc}}{\rho_{effc}}$ &$\rho_{effc}$ &$p_{reffc}$ &$p_{teffc}$ &$\rho_{c}$&$p_{c}$&$\rho^{D}_{c}$ &$p_{rc}^{D}$ &$p_{tc}^{D}$\\
\hline
$Her\;X-1$ & 0.85& 8.5  & $<1$ &$3.0884\times10^{-3}$ &$4.4052\times10^{-4}$ &$4.4052\times10^{-4}$&0.0027 &0.0007&0.00035&-0.00035 &-0.00035\\
\hline
 $LMC\;X-4$& 1.29&  --- & $<1$ &$5.3836\times10^{-3}$  &$9.3519\times10^{-4}$ &$9.3519\times10^{-4}$& 0.0047&0.0015&0.00061& -0.00061&-0.00061\\
\hline
 $Cen\;X-3$& 1.49 & --- &$<1$ & $6.6688\times10^{-3}$ & $1.2865\times10^{-3}$&$1.2865\times10^{-3}$& 0.0043&0.0014&0.00057& -0.00057&-0.00057\\
\hline
 $PSR\;J1614-2230$& 1.97  &--- &$<1$ &$1.0673\times10^{-2}$& $2.8107\times10^{-3}$&$2.8107\times10^{-3}$& 0.0040&0.0015&0.00052& -0.00052&-0.00052\\
\hline
 $PSR\;J0740+6620$& 2.14 & --- &$<1$ & $1.2528\times10^{-2}$ & $3.7932\times10^{-3}$&$3.7932\times10^{-3}$& 0.0040&0.0015&0.00052& -0.00052&-0.00052\\
\hline
 $GW190814$& 2.50 & --- &$<1$ & $1.7646\times10^{-2}$& $7.8450\times10^{-3}$&$7.8450\times10^{-3}$& 0.0036&0.0015&0.00047& -0.00047&-0.00047\\
\hline
\end{tabular}
\end{center}
\end{table*}

\section{Discussion and evaluation of calculated results}\label{sec4}
We investigate the DE stellar models in this section of our case study by selecting the appropriate functions $f(T)$ and $h(T)$, respectively, from the literature \cite{ditta2023effect,bamba2012dark}. We select the compact star model $Her\;X-1$ for the whole graphical analysis. We then compare the pressures and densities of DE compact stellar candidates, as shown in table-\ref{tab2} and \ref{2}, including $Her\;X-1,\;LMC\;X-4,\;Cen\;X-3,\;PSR\;J1614-2230,\;PSR\;J0740+6620\;\&\;GW190814$. The following is a synopsis of our acquired results:
\subsection{Energy densities profiles and pressures profiles of dark energy models for compact objects }

 Physical validity is crucial for studying compact object models because a study that is not physically admissible may not be worth doing. One tool used to guarantee the physical affirmation of the study is the density parameters and the pressure components. The behaviour for density and pressure patterns in compact designs with DE should be as follows:
 \begin{itemize}
     \item  In normal matter distribution density and pressure profiles should be positive: $\rho_{eff}>0,\;p_{reff}\geq0,\;p_{teff}>0,\;\rho>0\;\&\;p>0$.
     \item in DM distribution, the density should be positive, and pressures must be negative:  $\rho^{D}>0,\;p_{r}^{D}<0\;\&\;p_{t}^{D}<0$.
     \item Further, at the center, the above profiles should be Maximum in numeric values with the positive and negative signs whatever is applicable, and at the boundary, they should tend toward the zero; the mean should be minimum in numeric values, retaining their signs.
 \end{itemize}

  Considering the propagation of the effective pressures $p_{reff}\;\&\;p_{teff}$ and the effective energy parameter $\rho_{eff}$, Fig. (\ref{1}) contains useful details. As per the physical standards, the energy density and pressure components exhibit the highest values near the center (as also shown in Tables-\ref{tab1}-\ref{tab2}) and uniform, positive declines throughout the star from center towards the boundary ($0<r\leq R$). This validates the celestial body's physical legitimacy.

  Fig. (\ref{2}) compares the central effective densities and effective pressure of DE compact stars candidates $Her\;X-1,\;LMC\;X-4,\;Cen\;X-3,\;PSR\;J1614-2230,\;PSR\;J0740+6620\;\&\;GW190814$. This confirms that the star candidates with more masses have higher central effective densities and effective radial and tangential pressure. Detailed information regarding central densities and pressure for DM and normal distributions is provided in table- \ref{tab2}.

  Moreover, energy profile $\rho$ and pressure profile $p$ are shown in Fig. (\ref{3}), which also authenticate the physical acceptability of normal distribution as mentioned earlier in acceptability criteria.

  Moreover, our study fully admits the DM distribution criteria in Fig. (\ref{4}). One can easily extract information from these graphs as, $\rho^{D}>0,\;p_{r}^{D}<0,\;\&\;p_{t}^{D}<0$. This fact is also mentioned in Tables-\ref{tab1}-\ref{tab2} in the case of central values.
\subsection{Anisotropy profile, gradients profiles, and adiabatic index profiles}

Anisotropy, the repulsive force provided by $\Delta=p_{teff}-p_{reff}$ neutralizes the consequences of gradient components, significantly improving the equilibrium balance  and stability of stellar configurations. The observable positive anisotropic conduct attests to the long-lasting advantages of the said repulsive forces. The requirement that $\Delta>0$ within the boundary of the compact body for $p_t>p_r$, where determines the anisotropy. Nevertheless, $\Delta\rightarrow0$ as $r\rightarrow0$, where $p_{reff}=p_{teff}$. The anisotropy $\Delta$ illustrated in Fig. (\ref{5})'s first graph complies with the necessary behavior.

Gradients usually have a $0$ value at the center and a negative, declining trend. The expression $(\frac{d\rho_{eff}}{dr}=\frac{d p_{reff}}{dr}=\frac{d p_{teff}}{dr}|{r\rightarrow0}=0)$, with the exception of the following in their graphical portrayal: $(\frac{d\rho_{eff}}{dr},\;\frac{d p_{reff}}{dr},\;\frac{d p_{teff}}{dr})|{0<r\leq R}<0$. The anticipated gradients fall within this range, as shown by the results in the second graph of Fig. (\ref{5}).

Chandrasekhar investigated stability under the adiabatic index in \cite{chandrasekhar1964dynamical}. The adiabatic index stability constraint was established by Heintzmann and Hillebrandt~\cite{heintzmann1975neutron} by resolving the inequality $\Gamma \Big|_{0 \leq r \leq R} = \frac{4}{3}$. The adiabatic index can be expressed mathematically as follows:
\begin{eqnarray}
    \Gamma = \frac{p_{reff} + \rho_{eff}}{p_{reff}}v^2_r. \label{adiabatic}
\end{eqnarray}

The third graph of Fig. (\ref{5}) shows the graphical behavior of the adiabatic index. It is important to note that in a stable polytropic celestial body, the adiabatic index must be larger than $\frac{4}{3}$, by a factor that relies on the ratio $\frac{\rho}{p_r}$ at the core of the compact heavenly object~\cite{glass1983stability}. According to Harrison (1965), EoS associated with neutron star matter has a ratio $\frac{\rho}{p_r}$ that ranges from two to four. In this regard, the graphical analysis shows that the adiabatic index is always bigger than $\frac{4}{3}$ throughout the heavenly distribution; as a result, our celestial object model admits a stable configuration concerning the adiabatic index.
\subsection{Energy limits profiles}

The EMT in GR describes mass, momentum, and stress and depicts the arrangement of gravitation-free fields (GFF) and matter fields in space-time. Nevertheless, neither the admissible GFF in the space-time manifold nor the state of matter are directly related to the Einstein field equations (EFEs). Energy requirements, which validate all sorts of fluid, oppose GFF in GR, and guarantee a feasible and physically admissible allocation of matter, are used to establish physically legitimate solutions of the field equations. To attain this distribution, the anisotropic behavior of energy should stay positive and adhere to certain constraining limits across the star body. Eqs (\ref{48}-\ref{51}) express these limitations, these limits are well famed as the Strong Energy Condition (SEC), Weak Energy Condition (WEC), Null Energy Condition (NEC), and the Dominant Energy Condition (DEC):
\begin{eqnarray}
\mathrm{SE}C&:&\rho_{eff}+p_{\gamma\{eff\}\}}\geq0,\rho_{eff}+p_{reff}+2p_{teff}\geq0,\label{48}\\
\mathrm{WEC}&:&\rho_{eff}\geq0,\rho_{eff}+p_{\gamma\{eff\}}\geq0,\label{49}\\
\mathrm{NEC}&:&\rho_{eff}+p_{\gamma\{eff\}}\geq0,\label{50}\\
\mathrm{DEC}&:&\rho_{eff}>|p_{\gamma \{eff\}}|.\label{51}
\end{eqnarray}
 Here, $r$ \& $t$ stand for the radial cordinate if $\gamma = r$) and tangential coordinates if $\gamma = t$.  Our results, which are shown in Fig. (\ref{6}), are consistent with the standard parameters used in studies of compact stars. These conditions were violated in the case of dark energy matter, as shown in the last graph of Fig (\ref{6}).

 \subsection{Equation of state profiles}

Whether a compact star system is made of DM or normal matter, its makeup matters greatly. The value ranges of $w_r$ and $w_t$ for normal or Byronic matter EoS must lie between $0 \leq w_r < 1$ and $0 < w_t < 1$ to guarantee that the system is composed of normal matter. The phrases for EoS are provided by:
\begin{equation}\label{54}
w_r = \frac{p_{reff}}{\rho}=\;\;\;\;\& \;\;\;\;w_{teff} = \frac{p_{teff}}{\rho},
\end{equation}
while in the case of dark energy stars, these equations are:
\begin{equation}\label{54a}
w_r^D = \frac{p_{r}^D}{\rho}\;\;\;\;\& \;\;\;\;w_{t}^D = \frac{p_{t}^D}{\rho}.
\end{equation}
The first two graphs in the upper panel of Fig. (\ref{7}) demonstrate that these EoS parameters meet the necessary constraining conditions, guaranteeing that matter is dispersed uniformly realistic across the system. For the DM state, the EOS range should be  $w_r^D \;\&\; < w_t^D < -1/3$; this criterion is also justified in the upper panel graphs of Fig. (\ref{7}).

\subsection{Sound speeds}

This article will examine two stability parameters, $v_r^2$, the radial direction speed, and $v_t^2$, the tangential direction speed, to assess the stability of the stellar configurations. We also need to consider the notion of anisotropic fluid distribution, sometimes referred to as the Herrera Cracking idea. According to the Herrera Cracking idea, to preserve stability, the sound speeds need to meet the following conditions: $0 < v_r^2, v_t^2 < 1$; both speeds need to be smaller than the speed of light, and the speed of light is $c=1$. The following is the formula for sound speeds:
\begin{equation}\label{55}
v_r^2 = \frac{dp_{reff}}{d\rho_{eff}} \;\;\&\;\; v_t^2 = \frac{dp_{teff}}{d\rho_{eff}}.
\end{equation}
An additional stability criterion was presented by Abreu et al. \cite{abreu2007sound}, which states that if $v_r^2 > v_t^2$ and $v_r^2 - v_t^2$ remains unaltered, the region is deemed possibly stable. The condition was further generalized to $0 < |v_t^2 - v_r^2| < 1$ by Andreasson \cite{andreasson2008sharp}, meaning the negation of cracking depicting the stable zone. Our results are compatible with the Abreu and Andreasson stability criteria, suggesting that our results for compact star investigation are stable. These results are displayed in Fig. (\ref{8}). It is noteworthy that the sound speeds $v_r^2>1 \;\&\; v_t^2>1$ demonstrate the system's instability for $\kappa=2.0$. Our DE-based star system is stable for all other selected values of $\kappa$.

\subsection{Behaviors of mass profile, compactness profile, and redshift profile.}
One important indicator of a star's compactness is the $\frac{m(r)}{r}$ compactness ratio. The formula that follows can be used for determining the mass:
 \textcolor{black}{\begin{equation}\label{56}
m(r)=4\pi\int \left(r^2\rho\right)dr,
\end{equation}}
We may obtain formulas for the redshift function $z_s$ and the compactness parameter $u(r)$ through the inclusion of the effects of Eq. (\ref{56}):
\begin{eqnarray}
    u &= \frac{m(r)}{r},\label{57}\\
z_s &= e^{-\frac{\nu(r)}{2}}-1\label{58}.
\end{eqnarray}
Buchdahl \cite{buchdahl1959general} determined the highest value of $u=\frac{m(R)}{R}<\frac{4}{9}$ for the compactness criterion. In \cite{abreu2007sound}, this criterion was extended to anisotropic matter configurations. Additionally, Buchdahl established a maximum value criteria for $z_s\leq4.77$, the redshift parameter \cite{bowers1974anisotropic}. The third graph of Figs. (\ref{6}) displays the smooth and regular mass function that our investigation produced. The compactness and redshift parameters obtained from our research, meet the physical admissibility requirements for the star system, as shown by the first and second graphs in Fig. (\ref{6}).

\section{conclusion}\label{sec5}
The foundation of this research on compact bodies system is a simple modification of $f(T)$, namely the MRTG theory, that is distinct from both the Rastall theory and $f(T)$ gravity since it incorporates Rastall's term and uses a torsion-based function. We include the vacuum case of the Schwarzschild solution to serve as an outer solution and the Karmarkar ansatz based upon the spherical symmetric space-time as an inside solution to achieve results which are physically admissible. We note that the results are strongly affected by the Rastall parameter, as one can retrieve the $f(T)$ theory of gravity by setting the Rastall parameter equal to zero. We investigate MTRG for dark energy stellar structures by selecting the particular modifications of gravity based on the functions $f(T)$ and $h(T)$. To examine the anisotropic conduct of the DE field in detail, we thoroughly analyzed the results for compact stellar candidate $Her\;X-1$ for different values of metric constant $\kappa$ (see Figs. \ref{1}-\ref{9}). In particular, we also determine the central densities and pressures for candidates $Her\;X-1,\;LMC\;X-4,\;Cen\;X-3,\;PSR\;J1614-2230,\;PSR\;J0740+6620\;\&\;GW190814$ (see table-\ref{tab2} and Fig. (\ref{2})). Some Works on compact stellar modeling in Rastall-based modifications of torsional theory have previously been done in literature \cite{ditta2023anisotropic,ditta2023effect,ditta2022anisotropic,ashraf2023structural}. But this work is different from the previous literature, as we first time studied dark energy stellar models in different forms of MTRG functions $f(T)\;\&\;h(T)$. The following is a summary of our study's primary findings:

 We accurately analyzed all parameters for dark energy compact models to show their anisotropic behavior. Dark matter (DM) parameter $\rho^D$ has positive conduct, while both pressure components $p_r^D$ and $p_t^D$ are negative for DM. Our results for $\rho_{eff}$, $p_{reff}$, $p_{teff}$, $\rho$, and $p$ contain the positive conduct depicting the real nature of matter configurations. Furthermore, $\Delta$, the anisotropy parameter, exhibits smooth behavior with negative gradients from the center to the boundary. Energy conditions exhibit positive behavior throughout the stellar configurations, whereas they are negative in the case of dark energy density and pressures; negative EoS parameters also validate the behaviour of the DE; the speed of sound indicates stability for the chosen range of $\kappa$, except $\kappa=2.0$; and the causality limits satisfy the necessary conditions. The system is stable because the mass function, compactification, and redshift functions behave as they should according to the acceptable criteria.

 Our research shows that dark energy, with its constant energy density and negative pressure, plays a major role in the stellar structure of dark energy stars. This part changes the star's size, stability, and energy density distribution, which in turn changes the star's overall size and stability. The standard matter and energy present would govern the star's interior structure if dark energy did not contribute. Because of this, we would see different density and pressure profiles, which may cause the star's size and stability to vary. Dark energy is a repulsive force that, in the absence of it, could cause the star to shrink in size and change its stability properties. Models with and without dark energy \cite{ditta2023effect} were compared to quantify these changes. The estimated radius and mass of the star change noticeably when dark energy is excluded, according to our findings. These results highlight the need to incorporate dark energy into precise models of such small objects, and we recognize that dark energy is a key component in determining the star's characteristics.

 This research significantly unravels the interplay between modified Rastall teleparallel gravity and dark energy within compact stars. By investigating the celestial mystery of dark energy in the modified gravitational scenario, our study enhances the understanding of the cosmos. It motivates future studies in the modified theories of gravity.

\section*{Acknowledgement}

Xia Tiecheng acknowledge this paper to be funded by the National Natural Science Foundation of China 11975145. Asif Mahmood would like to acknowledge the Researcher's Supporting Project Number (RSP2025R43), King Saud University, Riyadh, Saudi Arabia,

\section*{Conflict Of Interest Statement }
The authors declare that they have no known competing financial interests or personal relationships that could have appeared to influence the work reported in this paper.

\section*{Data Availability Statement} This manuscript has no associated data, or the data will not be deposited.
(There is no observational data related to this article. The
necessary calculations and graphic discussion can be made available
on request.)

\end{document}